\documentclass[sigconf]{acmart}
\usepackage{algorithm}
\usepackage{algorithmic}
\usepackage{newfloat}
\usepackage{multirow}
\usepackage{tabularx}
\usepackage{graphicx}
\usepackage{caption}
\usepackage{subfigure}
\usepackage{xcolor}
\usepackage{listings}
\usepackage{color}
\usepackage{amsfonts}
\usepackage{bm}
\usepackage{stfloats}
\usepackage{svg}
\svgsetup{
    inkscapepath=i/svg-inkscape/
}
\svgpath{{figures/}}
\newcommand{\textbi}[1]{\textbf{\textit{#1}}}

\lstdefinestyle{lfonts}{
  basicstyle   = \small\ttfamily,
  stringstyle  = \color{purple},
  keywordstyle = \color{blue!60!black}\bfseries,
  commentstyle = \color{olive}\scshape,
}
\lstdefinestyle{lnumbers}{
}
\lstdefinestyle{llayout}{
  aboveskip        = 0mm,
  belowskip        = 0mm,
  breaklines       = true,
  tabsize          = 2,
  columns          = flexible,
}
\lstdefinestyle{lgeometry}{
  xleftmargin      = 10pt,
  xrightmargin     = 0pt,
  frame            = tb,
  framesep         = \fboxsep,
  framexleftmargin = 10pt,
}
\lstdefinestyle{lgeneral}{
  style = lfonts,
  style = lnumbers,
  style = llayout,
}
\lstdefinestyle{python}{
    language = {Python},
    style    = lgeneral,
}
\lstdefinestyle{java}{
    language = {Java},
    style    = lgeneral,
}

\AtBeginDocument{%
  \providecommand\BibTeX{{%
    \normalfont B\kern-0.5em{\scshape i\kern-0.25em b}\kern-0.8em\TeX}}}

\copyrightyear{2022}
\acmYear{2022}
\setcopyright{acmcopyright}\acmConference[ICPC '22]{30th International Conference on Program Comprehension}{May 16--17, 2022}{Virtual Event, USA}
\acmBooktitle{30th International Conference on Program Comprehension (ICPC '22), May 16--17, 2022, Virtual Event, USA}
\acmPrice{15.00}
\acmDOI{10.1145/3524610.3527903}
\acmISBN{978-1-4503-9298-3/22/05}

\begin{document}

\title{GypSum: Learning Hybrid Representations for Code Summarization}

\author{Yu Wang}
\email{wangyu@stu.ecnu.edu.cn}
\orcid{1234-5678-9012}
\affiliation{%
  \institution{East China Normal University}
  \city{Shanghai}
  \country{China}
 }

 \author{Yu Dong}
\email{yudong@stu.ecnu.edu.cn}
\affiliation{%
  \institution{East China Normal University}
  \city{Shanghai}
  \country{China}
 }

\author{Xuesong Lu}
\authornote{Corresponding author}
\email{xslu@dase.ecnu.edu.cn}
\affiliation{%
  \institution{East China Normal University}
  \city{Shanghai}
  \country{China}
 }

\author{Aoying Zhou}
 \email{ayzhou@dase.ecnu.edu.cn}
\affiliation{%
  \institution{East China Normal University}
  \city{Shanghai}
  \country{China}
 }

\renewcommand{\shortauthors}{Y Wang, Y Dong, X Lu and A Zhou.}

\begin{abstract}
Code summarization with deep learning has been widely studied in recent years. Current deep learning models for code summarization generally follow the principle in neural machine translation and adopt the encoder-decoder framework, where the encoder learns the semantic representations from source code and the decoder transforms the learnt representations into human-readable text that describes the functionality of code snippets. Despite they achieve the new state-of-the-art performance, we notice that current models often either generate less fluent summaries, or fail to capture the core functionality, since they usually focus on a single type of code representations. As such we propose GypSum, a new deep learning model that learns hybrid representations using graph attention neural networks and a pre-trained programming and natural language model. We introduce particular edges related to the control flow of a code snippet into the abstract syntax tree for graph construction, and design two encoders to learn from the graph and the token sequence of source code, respectively. We modify the encoder-decoder sublayer in the Transformer's decoder to fuse the representations and propose a dual-copy mechanism to facilitate summary generation. Experimental results demonstrate the superior performance of GypSum over existing code summarization models.
\end{abstract}




\maketitle

\section{Introduction}
\label{sec:introduction}
Source code documentation is particularly important in software engineering since it facilitates software development, bug fixing and software maintenance~\cite{de2005study,xia2017measuring,yang2021incbl}. However, writing documentation takes time and is usually postponed by the developers towards the end of the project, only if time permits. To help with the documentation task, recent works have investigated the possibility of automatically generating a piece of readable description for a function, which is referred to as the ``code summarization'' task, using deep learning techniques~\cite{ahmad-etal-2020-transformer,hu2018deep,hu2018summarizing,wan2018improving,wei2019code}.

Current deep learning models for code summarization usually adopt the encoder-decoder framework and vary the detailed structure of the two components. The encoder converts a code snippet into latent representations, based on which the decoder outputs a natural language summary describing the effect of the snippet. For example, inspired by Neural Machine Translation~\cite{bahdanau2015neural,sutskever2014sequence}, Hu et al.~\cite{hu2018deep} utilize the sequence-to-sequence architecture for code summarization, where the LSTM encoder learns the latent code representations from the serialized abstract syntax tree (AST) and the LSTM decoder produces the summary using the attention mechanism. Other works either design a particular encoder/decoder to leverage a specific input form of source code, i.e., the AST or token sequence~\cite{leclair2019neural,leclair2020improved,shi2021cast,wu2021code}, or propose specific learning objectives to facilitate model training~\cite{hu2018summarizing,wan2018improving,wei2019code}. In our study, we observe that the latent representations learned from different code forms may capture quite different semantic information, resulting in varying behaviors in the decoded summaries. For instance, Table~\ref{tab:motivatingExample} shows a function and the corresponding summaries generated by three models. The ``Transformer'' model~\cite{ahmad-etal-2020-transformer} uses the Transformer's encoder to learn from the text token sequence of a snippet, whereas the ``DeepCom'' model~\cite{hu2018deep} uses an LSTM encoder to learn from the flattened AST. We observe that the summary generated by Transformer is more detailed and fluent, but fails to describe the core effect of the function. It says the function adjusts \emph{the column width} while the ground-truth is adjusting \emph{the row heights}. Moreover, the text \emph{for the specified table and column names} makes little sense. On the other hand, although the summary generated by DeepCom describes the functionality closer to the ground-truth, it is too brief and grammatically incorrect.

Based on the observation, we conjecture that the representations learnt from token sequences preserve better the \emph{naturalness} of the language and can benefit text generation, whereas the representations learnt from AST structures capture better the functional information, thereby improving the \emph{informativeness} of the summary. If the conjecture is correct, learning simultaneously the representations from the token sequences and AST structures should benefit the generation of summaries that are both natural and informative.

This motivates us to propose ``GypSum'', a new deep model for code summarization, which leverages two encoders to learn hybrid latent representations from source code. Although several previous studies~\cite{haque2020improved,leclair2020improved,shi2021cast} have investigated the similar idea, we show that GypSum not only achieves the new state-of-the-art performance, but also is more interpretable when generating the summaries (See Section~\ref{sec:visual}). In the example of Table~\ref{tab:motivatingExample}, the summary generated by GypSum is fluent and correctly describes the functionality of the code.
\begin{table}[!t]

  \caption{A motivating example showing the varying behaviors of summaries generated using different representations.}
   \label{tab:motivatingExample}
      \begin{tabular}{p{0.45\textwidth}}
      \hline
    \begin{lstlisting}[style = java,basicstyle=\footnotesize]
private void adjustRowHeights(Jtable table){
  for(int row=num; row<table.getRowCount(); row++){
    int rowHeight = table.getRowHeight();
    for(int column=num; column<table.getColumnCount(); column++){
      component comp = table.prepareenderer(table.getCellenderer(row, column), row, column);
      rowHeight = math.max(rowHeight, comp.getPreferredSize().height);
    }
    table.setRowHeight(row, rowHeight);
  }
}
\end{lstlisting} 
\small\textcolor{green}{Ground-Truth:}  adjust the row heights of a table based on the table contents.\\
\small\textcolor{red}{GypSum:}  adjust the row heights of the table based on the preferred size.\\
\small\textcolor{orange}{Transformer:}  adjust the \textcolor{cyan}{column width} of the table for \textcolor{cyan}{the specified table and column names} .\\
\small\textcolor{purple}{DeepCom:} adjust \textcolor{cyan}{column height} based on contents.\\
      \hline
\end{tabular}
\end{table}
GypSum consists of two encoders and one decoder. The two encoders leverage graph attention neural networks~\cite{li2015gated} (GAT) and a pre-trained model for programming language and natural language~\cite{feng2020codebert} (PL-NL), respectively, to learn the latent representations from the AST-base graphs and token sequences of code snippets. The intuition is to use the attention mechanism in GAT to capture the key functional nodes in ASTs. The two types of representations are fused into hybrid representations in the decoder, and a dual-copy mechanism is proposed to facilitate summary generation. Besides the model design, our another contribution is the proposal of introducing semantic edges pertaining to control flows for graph construction, inspired by the control flow graph of a program.

In the experimental section, we show GypSum outperforms existing representative code summarization models by a fairly large margin on two commonly used datasets. We also conduct ablation study, case study and user study to justify the architecture choice and the performance of GypSum. Particularly, even without relying on any pre-trained PL-NL model, GypSum can still achieve the best performance compared with existing models.
\section{Related Work}
\label{sec:related_work}
Recent works on code summarization have mainly used the encoder-decoder framework. \citet{iyer2016summarizing} first use LSTM networks with attention to produce sentences that describe C\# code snippets and SQL queries. \citet{hu2018deep} propose a sequence-to-sequence model to learn from the ASTs. They propose a structure-based traversal method to flatten the ASTs into token sequences. Following this work, they propose to learn from API sequences extracted from source code, and use the learned API knowledge to enhance the performance~\cite{hu2018summarizing}. \citet{wei2019code} investigate the duality between the code generation task and the code summarization task. They use two sequence-to-sequence neural networks with attentions to train the two tasks simultaneously. \citet{ahmad-etal-2020-transformer} show that the quality of the summaries can be greatly improved by carefully implementing the Transformer structure. They replace the original positional encoding in the Transformer with the relative positional encoding, which encodes the pairwise relationship between the tokens in the token sequence. As an improvement to Transformer, \citet{gao2021code} propose to exploit code structural property and introduce constraints to the multi-head self-attention module. \citet{zhang2020retrieval} and \citet{wei2020retrieve} both propose a retrieval-based method, which improves summary generation by leveraging the comments of similar code snippets. \citet{shi2021cast} propose to split an AST into a set of subtrees and devise a recursive neural network to encode the subtrees. The encodings are then aggregated for generating the summary. \citet{choi2021learning} apply graph convolutions to obtain node representations from an AST and then input the sequence of node representations into the Transformer layers for code summarization. \citet{wu2021code} construct a multi-view adjacent matrix to represent the relationships between the tokens in source code, and use it to guide the self-attention computation in Transformer. Differently from above work, \citet{wan2018improving} discard the decoder structure and employ deep reinforcement learning networks for summary generation. They design an encoder consisting of an ordinary LSTM and a tree-based LSTM, which learn from serialized code snippets and binary trees transformed from ASTs. Other related work include~\cite{shido2019automatic,alon2018codeseq,leclair2020improved,chen2021novel}. Despite the great success, existing studies have not sufficiently explored the potentials of learning hybrid representations from the token sequence and the abstract syntax tree. We show that with the careful design of the model architecture, this simple idea can indeed outperform existing methods.
\section{The architecture of Gypsum}
\label{sec:approach}

\begin{figure*}[!htb]
  \centering
    \includegraphics[width=0.8\textwidth]{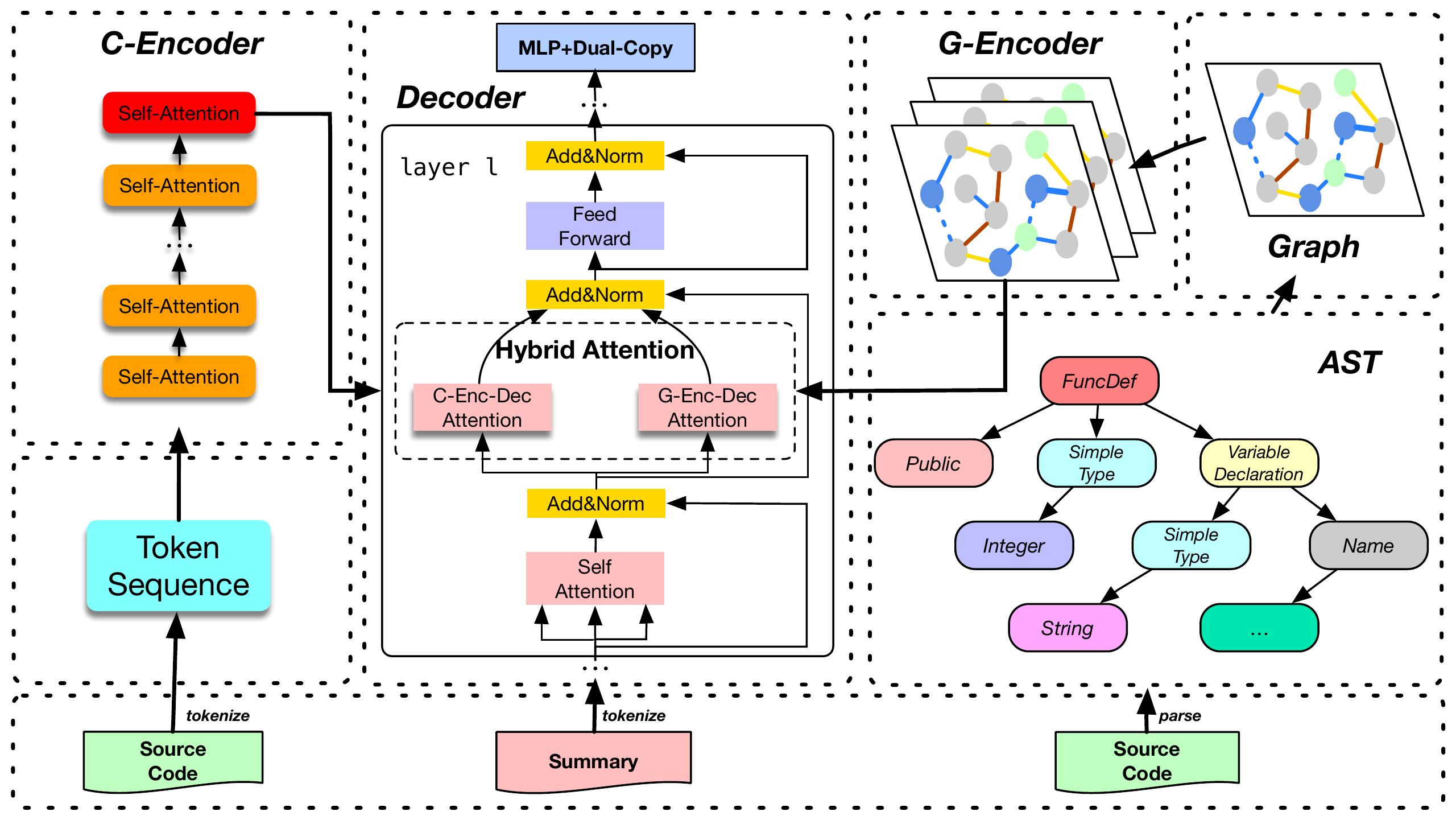}
  \caption{The architecture of the GypSum Model. The two encoders on the left and right part learn from token sequences and AST-based graphs, respectively. The decoder in the middle fuses the two types of learnt representations and outputs the summary.}
	\label{fig:GypsumArchitecture}
\end{figure*}

Gypsum consists of two encoders (\emph{c-encoder} and \emph{g-encoder}) and one decoder. The encodings are fused in the decoder for summary generation. We also propose a dual copy mechanism to enhance the token generation process. The overall architecture is depicted in Figure~\ref{fig:GypsumArchitecture}.

\subsection{\emph{c-encoder}: Encoding with Token Sequences}
The source code snippets preserve the human-injected naturalness of programming languages, i.e., they express certain semantics by following particular lexical and syntactic rules. Therefore we may simply convert a code snippet into a token sequence and learn the contextual representations of the tokens using NLP models. As our ultimate goal is to translate a code snippet into human-readable description, the representations should preserve the naturalness of programming language and also facilitate natural language generation. As such, a practical solution is to take the advantage of recently developed pre-trained models for programming and natural language (PL-NL). Two representative PL-NL pre-trained models are CodeBERT~\cite{feng2020codebert} and CodeT5~\cite{wang2021codet5}. We experimented with both and found CodeBERT produces slightly better results in our study. As such, we adopt CodeBERT to encode with the token sequences. It is worth noting that, experimental results in the ablation study (see Section~\ref{sec:ablation}) show our general solution outperforms existing methods even if we do not make use of any PL-NL pre-trained model.

CodeBERT is based on Roberta~\cite{liu2019roberta} and trained using a hybrid objective function that incorporates the task of masked language modeling (MLM)~\cite{devlin2019bert} and replaced token detection (RTD)~\cite{clark2019electra}. In the former task, the input is an individual function with the paired human-written documentation. During training, the input tokens are randomly masked and the task is to predict the masked tokens. In the latter task, the training phase employs a PL generator and an NL generator to recover the masked tokens in the function and the documentation~\footnote{In the task of RTD, the function and the documentation do not have to be paired.}, respectively, and trains CodeBERT as a discriminator to classify each token as `original' or `replaced'.

We obtain a pre-trained CodeBERT model from the official repository\footnote{https://github.com/microsoft/CodeBERT} and tune it on our datasets. We feed the tokens into CodeBERT to obtain the embeddings. We add an additional linear layer on top of its output layer and transform the embeddings so that they comply with the representations of the other encoder. Denoted by $l_c$, $d_{model}$, $d_e$ the length of the token sequence, the embedding size of each layer in CodeBERT and the final output embedding size of the encoder, the encoding process with CodeBERT can be formulated as follows,
\begin{equation}
  \tilde{\textbi{H}}_{c} = \textsc{CodeBERT}(\mathcal{C})
\end{equation}
\begin{equation}
  \textbi{H}_{c} = \textbi{W}_{c}\cdot\tilde{\textbi{H}}_{c},
\end{equation}
where $\mathcal{C}$ is the input sequence of the code tokens, $\tilde{\textbi{H}}_{c}\in \mathbb{R}^{d_{model}\times l_c}$ denotes the embedding matrix generated by CodeBERT,$\textbi{W}_{c}\in\mathbb{R}^{d_{e} \times d_{model}}$ denotes the parameters of the linear layer, and $\textbi{H}_{c}\in \mathbb{R}^{d_{e}\times l_c}$ denotes the final hidden representations. We refer to this encoder as \emph{c-encoder}, which is depicted in the left part of Figure~\ref{fig:GypsumArchitecture}.

\subsection{\emph{g-encoder}: Encoding with Graphs}
\label{sec:g_encoder}
The intermediate representations (IR) such as abstract syntax tree (AST) of programs are regularly used for code representation learning. These IRs are proven to reflect common syntactic and semantic patterns of code and lower the training cost~\cite{raychev2015predicting,alon2018general,alon2019code2vec}. There exist diverse approaches to leverage the ASTs, among which an effective way to further enrich the semantic structure of an AST is adding semantic edges between the nodes~\cite{allamanis2018learning,cvitkovic2019open}. Inspired by the control flow graph (CFG) of a program, we propose a new type of semantic edges for connecting AST nodes, namely \texttt{Control-Edge}, which reflect the flow of control statements such as \textit{if-else} and \textit{while}. Furthermore, we propose to use graph attention neural networks~\cite{velivckovic2018graph} to learn from the constructed semantic graphs, in order to better capture the key elements in the graphs using attentions. We refer to this encoder as \emph{g-encoder}, which is depicted in the right part of Figure~\ref{fig:GypsumArchitecture}.

\subsubsection{Extending the ASTs} 
\label{sec:extend_ast}
\begin{figure}[!htbp]
\centering
\subfigure[A snippet of Java code from a novice programmer.]{
\begin{minipage}[b]{0.3\textwidth}
\includegraphics[width=1\textwidth]{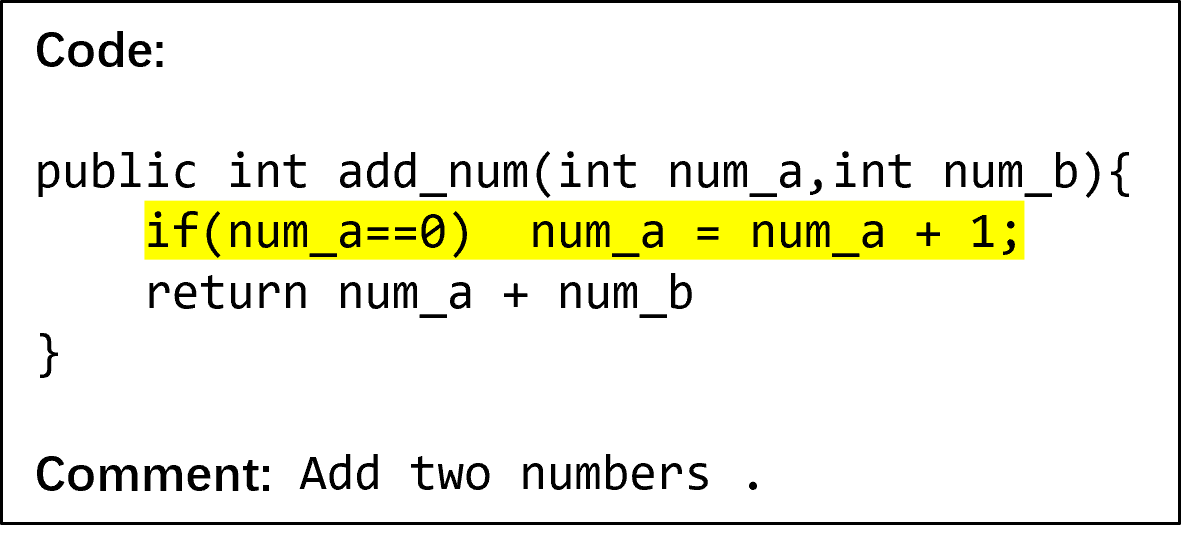}
\label{code_fig}
\end{minipage}
}
\subfigure[The full AST with induced edges of the above code, where the sub-tree in dashed red rectangle is shown in Figure~\ref{fig:partial_ast}]{
\begin{minipage}[b]{0.43\textwidth}
\includegraphics[width=1\textwidth]{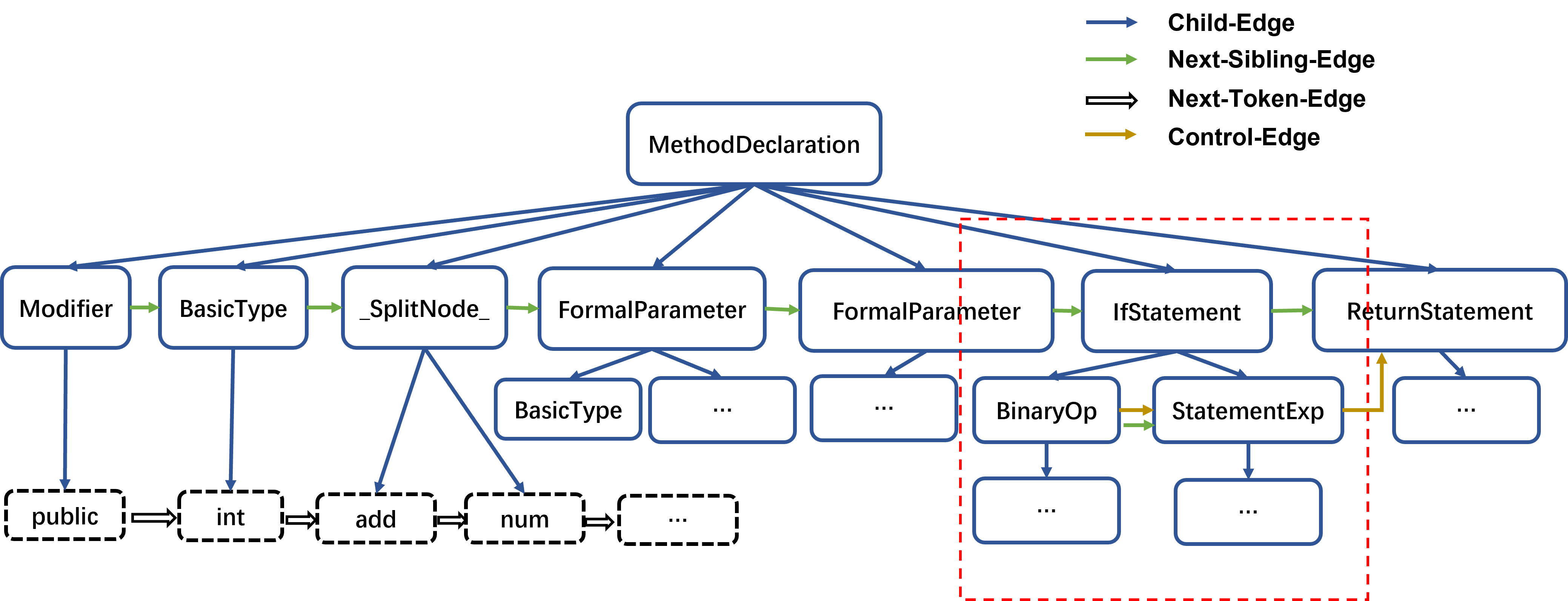}
\label{fig:example_ast}
\end{minipage}
}
\subfigure[Simplified semantic graph for line 2 of the above code, corresponding to the sub-tree in the red rectangle in Figure~\ref{fig:example_ast}]{
\begin{minipage}[b]{0.43\textwidth}
\includegraphics[width=1\textwidth]{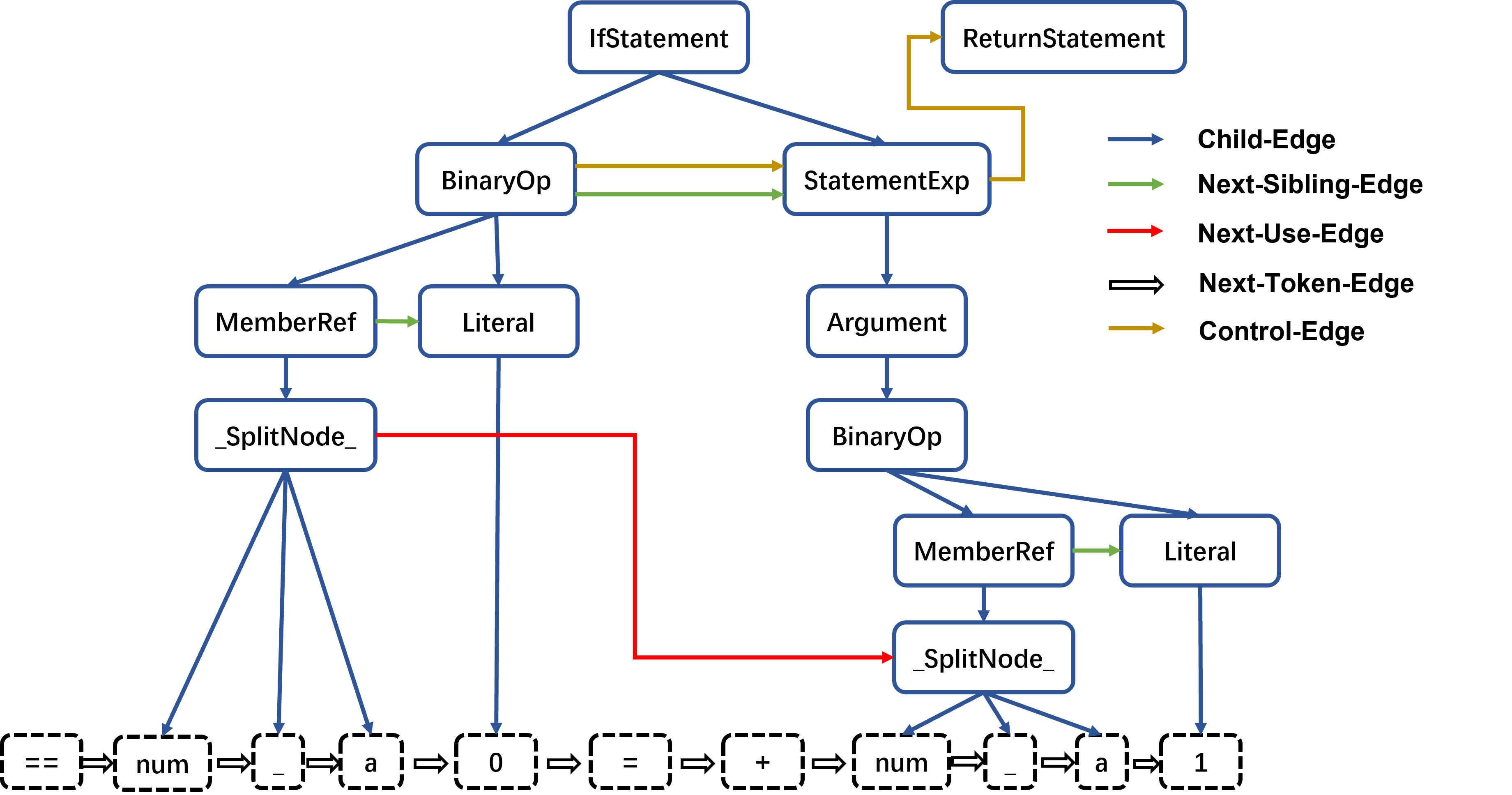}
\label{fig:partial_ast}
\end{minipage}
}
\caption{An example code snippet and its semantic graph.} \label{fig:example_graph}
\end{figure}

To better illustrate how we construct the semantic graphs, we use the code snippet in Figure~\ref{code_fig} as the running example. To generate the ASTs from source code, we use \textit{javalang}\footnote{https://github.com/c2nes/javalang} for Java code and use \textit{asttokens}\footnote{https://github.com/gristlabs/asttokens} plus \textit{ast}\footnote{https://github.com/att/ast} for Python code. In the constructed ASTs, the internal nodes are labeled with their class name such as \textsc{MethodDeclaration} and \textsc{StatementExp}, and the leaf nodes are labeled with the text of their attributes which correspond to the tokens in the source code.

We conduct two extensions to the ASTs generated by the tools. First, in order to share the vocabulary with c-encoder, we use the same tokenizer for the two encoders. As such, for both original Java and Python ASTs, the text in some leaf nodes are further tokenized into sub-words. For instance, `num\_a' in the example code in Figure~\ref{code_fig} is tokenized into `num',`\_' and `a'. For such a leaf node, we replace it with a special node \textsc{\_SplitNode\_} and make the sub-words as its children, forming the new leaf nodes. Second, for Python ASTs, some leaf nodes generated by the tools lose important attributes such as `args' and `name', which correspond to variables and function names, respectively. To mitigate the problem, we manually fetch the text of some important attributes and compose the additional leaf nodes to augment the original Python ASTs.

\begin{figure}[!htb]
    \includegraphics[width=0.43\textwidth]{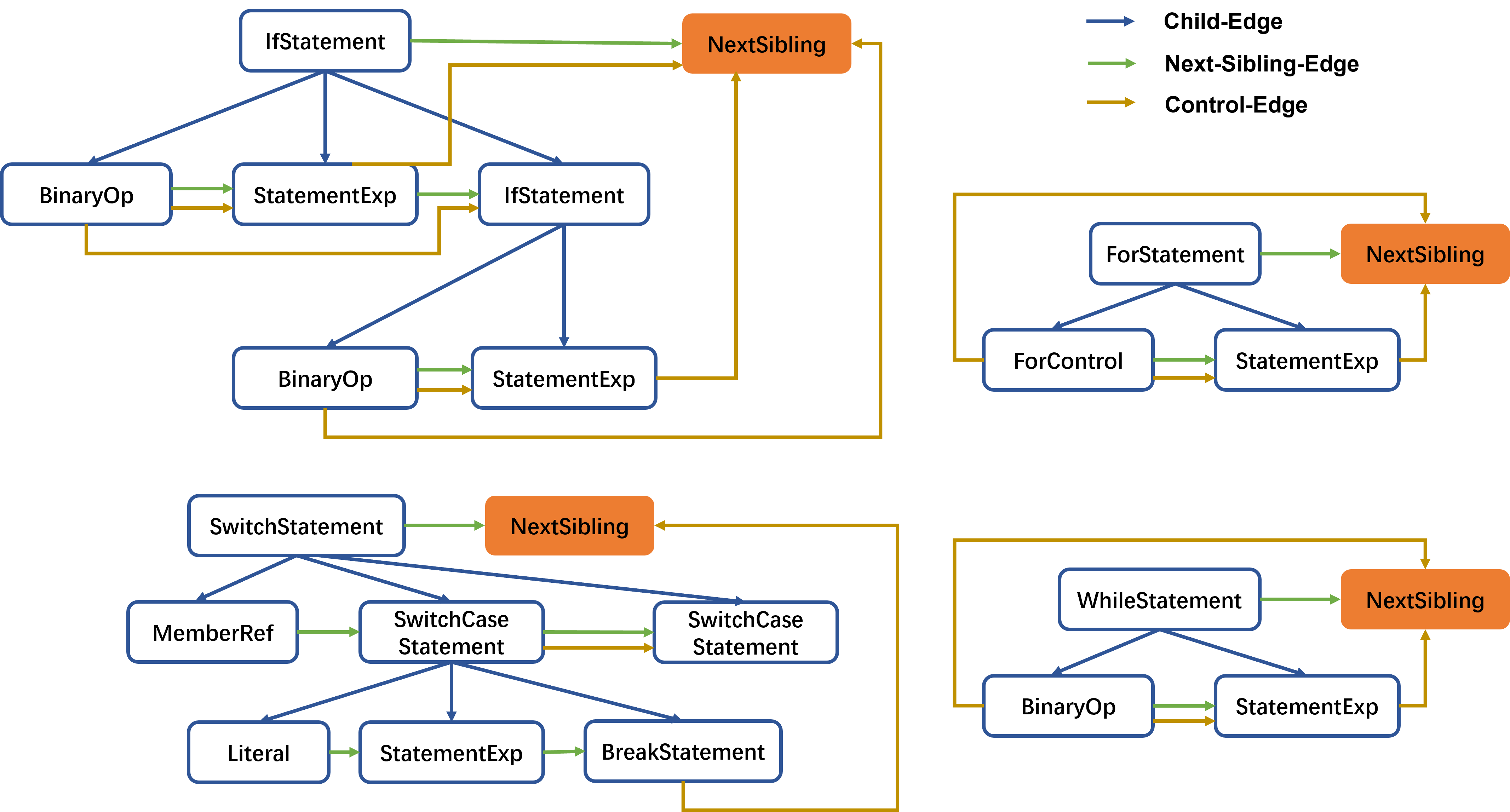}
    \caption{The Control-Edges for four types of control statements.}
	\label{fig:control_graph}
\end{figure}

\subsubsection{Graph Construction}\label{sec:graph_construction}
Following~\cite{allamanis2018learning,brockschmidt2018generative,schrouff2019inferring}, we first introduce four commonly used types of semantic edges into ASTs, namely, \texttt{Child-Edge}, \texttt{Next-Sibling-Edge}, \texttt{Next-Use-Edge} and \texttt{Next-To} \texttt{ken-Edge}. \texttt{Child-Edge}s are the original edges in ASTs plus the new edges obtained because of node splitting and augmentation. \texttt{Next-Sibling-Edge}s connect the sibling nodes in ASTs from left to right. \texttt{Next-Use-Edge}s connect a variable node to the next node of the same variable with depth-first search. Lastly, \texttt{Next-Token-Edge}s connect the leaf nodes from left to right so that the corresponding tokens comply with the order that they appear in the source code. The edges can be observed in Figure~\ref{fig:example_ast} and Figure~\ref{fig:partial_ast}, where Figure~\ref{fig:partial_ast} shows the detailed structure of the subtree inside the dotted rectangle of Figure~\ref{fig:example_ast}.

In addition to the above edges, we borrow the idea from the control flow graph and propose to introduce the \texttt{Control-Edge}s into ASTs, so that the data and control flow of a program could be better captured. In Figure~\ref{fig:control_graph}, we show the AST subtrees of four common control statements, i.e., \textit{if-else}, \textit{switch}, \textit{while} and \textit{for}. Take example of the \textit{if-else} subtree in the top-left corner, \textsc{BinaryOp}, \textsc{StatementExp} and \textsc{IfStatement} correspond to the \textit{if} condition, the statements in the main branch, and the \textit{else} branch. Since there can be flows from \textsc{BinaryOp} to \textsc{StatementExp} (i.e., the \textit{if} condition is true) and from \textsc{BinaryOp} to \textsc{IfStatement} (i.e., the \textit{if} condition is false), we initiate a \texttt{Control-Edge} from \textsc{BinaryOp} to \textsc{StatementExp} and from \textsc{BinaryOp} to \textsc{IfStatement}, respectively. Also, once the statements in \textsc{StatementExp} are executed, the flow should enter the next statement outside the \textit{if} block. As such, we initiate a \texttt{Control-Edge} from \textsc{StatementExp} to the next sibling (The \textsc{NextSibling} node stands for any possible sibling node.) of its parent node. The \texttt{Control-Edge}s for the other types of control statements are constructed using the similar principle.

For all types of edges, we introduce their corresponding reversed edges to enrich the semantic. Note that there may be multiple types of edges between two nodes. In such a case, we make use of them all in the computation of the model.

\subsubsection{Encoding with Graph Attention Neural Networks}
The Graph Attention Neural Networks~\cite{velivckovic2018graph} (GAT) recurrently update the state of a node by aggregating its own state and the states of all its neighbours at each step, using the attention mechanism. Denote by $\textbi{e}_i$ and $\textbi{e}_{ij}$, the embedding of node $i$ and the embedding of the edge from node $j$ to node $i$. GAT updates the node states as follows:
\begin{equation}
  \textbi{h}_i^{(0)} = \textbi{e}_i
\end{equation}
\begin{equation}
  s_{ij}^{(l)} = \textsc{LeakyReLU}(\bm{W}^{(l)}\cdot(\bm{W}\textbi{h}_i^{(l)}\oplus\bm{W}\textbi{h}_j^{(l)}\oplus\bm{W}_e\textbi{e}_{ij}))
\end{equation}
\begin{equation}
   a_{ij}^{(l)} = \frac{exp(s_{ij}^{(l)})}{\sum_{k\in \mathcal{N}(i)}exp(s_{ik}^{(l)})}
\end{equation}
\begin{equation}
  \textbi{h}_i^{(l+1)} = \sum_{j \in \mathcal{N}(i)}\sigma(a_{ij}^{(l)}\cdot\textbi{h}_j^{(l)}),
\end{equation}
where $\textbi{h}_i^{(l)}$ denotes the state of node $i$ at layer $l$, $\bm{W}$ and $\bm{W}_e$ are the shared weight matrices for nodes and edges, $\bm{W}^{(l)}$ is the weight matrix at layer $l$, $a_{ij}^{(l)}$ denotes the attention weight of contribution from node $j$ to $i$ at layer $l$. $\mathcal{N}(i)$ denotes the neighour nodes of node $i$.  

We employ multi-head for the attention computation. The node states of the last GAT layer compose the output matrix $\textbi{H}_{g}$ of g-encoder.

\subsection{The Fusion Decoder}
To fuse the two types of representations output by the two encoders, we design a fusion mechanism in the decoder to combine them and obtain the hybrid latent representations for decoding. We use the Transformer's decoder as the basic structure and retain the original hyperparameter settings. However, in the encoder-decoder attention sublayer, we apply the attentions of the decoder's hidden states to the output of each encoder, respectively, and then concatenate the output states of the two attention modules for further process. This is depicted in the ``hybrid attention'' module in Figure~\ref{fig:GypsumArchitecture}. In particular, given the output embeddings $\textbi{H}_{c}$ of the c-encoder and $\textbi{H}_{g}$ of the g-encoder and let $\textbi{D}^l_t = \{ \textbi{d}^l_1,\textbi{d}^l_2,\dots,\textbi{d}^l_{t-1}\}$ denote the hidden states of the decoder's $l^{th}$ layer before step $t$ (note that $\textbi{D}^0$ are the embeddings of the target summary), we can formulate the decoding process at step $t$ of layer $l$ as follows, 
\begin{equation}
  \tilde{\textbi{d}}^l_t = \textsc{attn}_s(\textbi{d}^{l-1}_t,\textbi{D}^{l-1}_{t+1},\textbi{D}^{l-1}_{t+1}),
\end{equation}
\begin{equation}
\textbi{attn}^l_t=[\textsc{attn}_c(\tilde{\textbi{d}^l_t},\textbi{H}_{c},\textbi{H}_{c}),\textsc{attn}_g(\tilde{\textbi{d}^l_t},\textbi{H}_{g},\textbi{H}_{g})],
\end{equation}
\begin{equation}
  \textbi{d}^l_t = \textbi{W}^{ff}\cdot\textsc{tanh}(\textbi{attn}^l_t),
\end{equation}
where $\textsc{attn}_s(\cdot)$ denotes the self-attention sublayer, and $\textsc{attn}_c(\cdot)$ and $\textsc{attn}_g(\cdot)$ denote the two attention modules in the encoder-decoder sublayer. The output of the two attention modules are concatenated and activated using Tanh. Finally, $\textbi{W}^{ff}$ are the parameters of the feed forward sublayer and $\textbi{d}^l_t$ is the encoding at position $t$ of layer $l$.

\subsection{The Dual-Copy Mechanism}
To better improve the ability of capturing keywords and solving the out-of-vocabulary problem, we propose a dual-copy mechanism to copy tokens from the two encoders. For the final hidden vector $\textbi{d}_t\in\mathbb{R}^{d_e}$ at position $t$ output by the decoder, we first compute the probability vector for choosing generation or copying as follows:
\begin{equation}
[p_v, p_c, p_g] = \textsc{Softmax}(\textbi{W}_{gen}\cdot \textbi{d}_t),
\end{equation}
where $p_v$, $p_c$ and $p_g$ are the probability for sampling a token from the vocabulary, copying from the input tokens of the c-encoder, and copying from the leaf nodes of the g-encoder, respectively. $\textbi{W}_{gen}\in\mathbb{R}^{3\times d_e}$ denotes the parameter matrix. $\textsc{Softmax}$ denotes the softmax function.

Then we compute the sampling distribution for generation and copying. For sampling from the vocabulary, the probability distribution is directly transformed from $\textbi{d}_t$ as follows:
\begin{equation}
  \textbi{p}_{voc} = \textsc{Softmax}(\textbi{W}_{voc}\cdot \textbi{d}_t),
\end{equation}
where $\textbi{W}_{voc}\in\mathbb{R}^{|vocab|\times d_e}$ denotes the parameter matrix. For copying, we can naturally use the last layer's attention weights of $\textbi{d}_t$ to all the encodings produced by each encoder, respectively, as the copy distribution, which can be formulated as follows:
\begin{equation}
\begin{split}
  \textbi{a}_{c} &= \textsc{Attn\_Score}_{c}^t, \\
   \textbi{a}_{g} &= \textsc{Attn\_Score}_{g}^t,
\end{split}
\end{equation}
where $\textbi{a}_{c}$ and $\textbi{a}_{g}$ are the probability distribution of copying from the input source code tokens and the leaf nodes, respectively. $\textsc{Attn\_}$ $\textsc{Score}_{c}^t$ and $\textsc{Attn\_Score}_{g}^t$ denote the attention weights of $\textbi{d}_t$ to all the encodings of the c-encoder and the g-encoder, respectively, in the last decoder layer.

Finally, the probability of producing a token $y_t$ in the summary at position $t$ is calculated as the joint probability:
\begin{equation}
 p(y_t) = p_v*p_{voc}(y_t) + p_c*\sum_{c_i=y_t}a_c^i + p_g*\sum_{g_i=y_t}a_g^i,
\end{equation}
where $c_i$ and $g_i$ denote the $i^{th}$ source code token and the $i^{th}$ leaf node, respectively. $a_c^i$ denotes the copy probability of the $i^{th}$ token.

\subsection{The Objective Function}
We use the negative log-likelihood as the objective function and attempt to minimize it during model training. Given a source code snippet $X_i$ and target summary $Y_i=\{y_1^i,y_2^i,\dots,y_T^i\}$, the objective function can be formulated as follows,
\begin{equation}
  \mathcal{L} = -\frac{1}{N} \sum^{N}_{i=1}\sum^{T}_{t=1}{log[p(y_t^i)]}
\end{equation}
where $N$ is the number of data points in the training data, $T$ is the length of the target summary.
\section{Performance Evaluation}
In this section, we evaluate GypSum and existing representative models for code summarization. We implement GypSum using Pytorch 1.7.0 and run all the experiments on a Tesla P40 GPU. All source code is available at \textcolor{blue}{\url{https://github.com/ICPC2022-Gypsum/GypSum_Code}}.

\subsection{The Datasets and Evaluation Metrics}\label{sec:dataset}
We use two frequently used public datasets in the experiments. One is a Java dataset~\cite{hu2018deep} containing $87,136$ Java code snippets with comments written by the developers. The other is a Python dataset~\cite{wan2018improving,hu2018deep} containing $87,226$ Python code snippets with comments. For fair comparison, we use the divided datasets in previous work \cite{ahmad-etal-2020-transformer,gao2021code}, where the proportions of training, validation and testing set are 8:1:1 and 6:2:2, respectively, for the Java and Python dataset.

We adopt three commonly used metrics for evaluating the performance of code summarization models, namely, BLEU~\cite{papineni2002bleu}, METEOR~\cite{banerjee2005meteor} and ROUGE-L~\cite{lin-2004-rouge}. The score of each metric on a testing set is computed as the average score of all the generated summaries. For all the metrics, a higher score indicates better performance.

\subsection{The Comparative Models}
We compare GypSum with eleven representative models. Except Rencos, CodeT5 and CodeBERT, we cite the numbers in their original papers since they use the same datasets as ours. We train Rencos, CodeT5 and CodeBERT on our datasets because they use different datasets in the original papers.

\begin{itemize}
  \item \textbf{RL+HybridSeq}~\cite{wan2018improving}. The model uses the LSTM-based encoder to learn from code snippets and binary trees transformed from ASTs. Then it uses the output of the encoder as the initial state and adopts reinforcement learning for code summary generation.
  \item \textbf{DeepCom}~\cite{hu2018deep}. The model uses an LSTM-based encoder-decoder architecture to learn from the flattened ASTs. A structure-based traversal method is proposed to traverse an AST and convert it into a sequence of tokens.
  \item \textbf{API+CODE}~\cite{hu2018summarizing}. The model learns the API knowledge from API sequences extracted from source code, and uses the learned API features to enhance the performance of code summarization.
  \item \textbf{Dual Model}~\cite{wei2019code}. The model investigates the duality between the code generation task and the code summarization task, and uses two sequence-to-sequence networks with attentions to train the two tasks simultaneously.
  \item \textbf{Transformer}~\cite{ahmad-etal-2020-transformer}. The model takes the Transformer structure and replaces the original positional encoding with the relative positional encoding, which encodes the pairwise relationships between the tokens in the source code text.
  \item \textbf{Rencos}~\cite{zhang2020retrieval}. The model first trains an attention-based encoder-decoder network using code and summaries. Then at inference time, it encodes a code snippet as well as two other most similar ones and fuses the encodings for code summary.
  \item \textbf{CAST}~\cite{shi2021cast}. The model hierarchically splits an AST into a set of subtrees and devises a recursive neural network to encode the subtrees. The encodings are then aggregated for generating the summary.
  \item \textbf{mAST+GCN}~\cite{choi2021learning}. The model applies graph convolutions to obtain node representations from an AST and then inputs the sequence of node representations into the Transformer layers for code summarization.
  \item \textbf{SiT}~\cite{wu2021code}. The model constructs a multi-view adjacent matrix to represent the relationships between the tokens in source code, and uses it to guide the self-attention computation in Transformer.
  \item \textbf{CodeT5}~\cite{wang2021codet5}. It is a pre-trained encoder-decoder model based on T5~\cite{raffel2020exploring} for programming and natural Language. We directly tune it with our datasets for code summarization.
  \item \textbf{CodeBERT}~\cite{feng2020codebert}. It is a pre-trained encoder model based on Roberta~\cite{liu2019roberta} for programming and natural Language. We add the Transformer decoder and tune the entire model with our datasets.
\end{itemize}

\subsection{The Hyperparameter Setting}
We report the hyperparameters of GypSum used in our experiments in Table~\ref{tab:hyper}. The output embedding size of the two encoders $d_e$ is set to 768 and the size of the node type embedding used in the graph node initialization layer $d_t$ is set to 128. For the c-encoder, the length of input tokens $l_c$ is truncated to 400 and 300 for the Java and Python dataset, respectively. The number of self-attention layers in the c-encoder $L_c$ is set to 12 and the embedding size of each layer $d_{model}$ is 768. In each layer, the number of heads $head_c$ is 12 and the sizes of the key and value are both set to 64. Finally, the dimension of the feedforward layer $d_{ff}$ is 2048. For the g-encoder, the number of output node embeddings $l_g$ is set to 300, the number of graph attention heads $head_g$ is set to 8, and the number of graph attention layers $L_g$ is set to 4. For the decoder, the number of layers $L_d$ is set to 6. The length of the input summary $l_s$ in training is truncated or zero-padded into 100 and 80 for the Java and Python datasets, respectively, in terms of the number of tokens.

When training, we set the dropout rate at the embedding and fully-connected layers to 0.2, learning rate to 0.0001, and batch size to 32, respectively. We use Adam for optimization. When testing, we use beam search for token generation and set the beam size to 6.
\begin{table}[h!]
\centering
  \caption{Hyperparameter settings for GypSum.}
  \label{tab:hyper}
  \begin{tabular}{c|c|cc}
    \hline
    & hyperparameter & Java & Python \\
    \hline
    embedding&$d_e$&\multicolumn{2}{c}{768}\\
    \hline
    \multirow{6}{*}{c-encoder}
    & $L_c$ & \multicolumn{2}{c}{12} \\
                & $l_c$ &  400 & 300 \\
        & $head_c$ & \multicolumn{2}{c}{12} \\
        & $d_{model}$ & \multicolumn{2}{c}{768} \\
        & $d_k$,$d_v$ & \multicolumn{2}{c}{64} \\
        & $d_{ff}$ & \multicolumn{2}{c}{2048} \\
    \hline
    \multirow{4}{*}{g-encoder}
            & $l_g$ & \multicolumn{2}{c}{300} \\
            & $h_g$ & \multicolumn{2}{c}{768} \\
            & $head_g$ & \multicolumn{2}{c}{8} \\
            & $L_g$ & \multicolumn{2}{c}{4} \\
    \hline
       \multirow{2}{*}{decoder}
            & $l_s$ & 100 & 80 \\
            & $L_d$ & \multicolumn{2}{c}{6} \\
    \hline
    \multirow{4}{*}{training}
                & dropout & \multicolumn{2}{c}{0.2} \\
                & optimizer & \multicolumn{2}{c}{Adam} \\
                & learning rate & \multicolumn{2}{c}{0.0001} \\
                & batch size & \multicolumn{2}{c}{32} \\
                    \hline
    testing& beam size & \multicolumn{2}{c}{6} \\
    \hline
    \end{tabular}
\end{table}

\subsection{Comparative Study}
\label{sec:experiment}
\begin{table*}
  \caption{The comparative results on the two testing sets.}
  \label{tab:full}
  \begin{tabular}{l|c|c|c|c|c|c}
    \hline
    \multirow{2}{*}{Methods}&\multicolumn{3}{c|}{Java}&\multicolumn{3}{c}{Python}\\
    \cline{2-7}
    &BLEU&METEOR&ROUGE-L&BLEU&METEOR&ROUGE-L\\
        \hline
    RL+HybridSeq (2018) \cite{wan2018improving}&38.22&22.75&51.91&19.28&9.75&39.24\\
    DeepCom (2018)\cite{hu2018deep}&39.75&23.06&52.67&20.78&9.98&37.35\\
     API+CODE (2018)\cite{hu2018summarizing}&41.31&23.73&52.25&15.36&8.57&33.65\\
     Dual (2019)\cite{wei2019code}&42.39&25.77&53.61&21.80&11.14&39.45\\
    Rencos (2020)\cite{zhang2020retrieval}&44.77&25.82&54.71&29.20&19.51&44.32\\
    Transformer (2020) \cite{ahmad-etal-2020-transformer}&44.58&26.43&54.76&32.52&19.77&46.73\\
    CAST (2021)
    \cite{shi2021cast}&45.19&27.88&55.08&-&-&-\\
    mAST+GCN (2021)
    \cite{choi2021learning}&45.49&27.17&54.82&32.82&20.12&46.81\\
    SiT (2021)
    \cite{wu2021code}&45.19&27.52&55.87&34.31&$\mathit{22.09}$&49.71\\
    CodeT5 (2021)
    \cite{wang2021codet5}&46.01&28.55&$\mathit{56.49}$&34.31&21.74&49.25\\
    CodeBERT (2020)
    \cite{feng2020codebert}&$\mathit{46.64}$&$\mathit{28.84}$&56.23&$\mathit{34.34}$&21.99&$\mathit{49.73}$\\
    \hline
    \multirow{2}{*}{\textbf{GypSum} (ours)}&{\bf 48.57}&{\bf 30.85}&{\bf 59.42}&{\bf 35.88}&{\bf 22.96}& \textbf{50.27}\\
    &($\uparrow$ 1.93)&($\uparrow$ 2.01)&($\uparrow$ 2.93)&($\uparrow$ 1.54)&($\uparrow$ 0.87)&($\uparrow$ 0.54)\\
    \hline
    \multicolumn{7}{c}{Ablation Study}\\
    \hline
    LSTM+g-encoder &46.01&28.87&56.39&33.82&21.19&47.85\\
    Transformer+g-encoder &47.31&30.02&58.12&35.01&22.06&49.73\\
    c-encoder+Code2Seq & 47.83&30.29&58.29&35.13&22.53&50.08\\
    c-encoder+ASTNN & 47.87&30.36&58.40&35.29&22.67&50.13\\
    c-encoder+GGNN &48.02&30.50&59.01&35.47&22.72&50.18\\
    \hline
    \end{tabular}
\end{table*}

\subsubsection{Main Results}
We first report the main results on the two testing sets in Table~\ref{tab:full}. The best result and the second best result for each metric (i.e., each column) are in the bold and italic font, respectively. CAST has not reported the results on the Python dataset in the original paper.

We observe that the GypSum model outperforms all the comparative models (shown in the top half of the table) by a notable margin for all the metrics on both datasets. Compared with the second best results, GypSum improves the BLEU, METEOR and ROUGE-L scores by 1.93, 2.01 and 2.93, respectively, on the Java dataset. The corresponding improvements on the Python dataset are 1.54, 0.87 and 0.54, respectively. The results prove the effectiveness of our idea for learning hybrid representations for source code summarization. Comparing between the results on the two datasets, we could observe that both the absolute values and the performance gains of the Python dataset are smaller than that of the Java dataset. This is probably because the original ASTs of Python code are much simpler than that of Java code, as we discussed in Section~\ref{sec:extend_ast}. Therefore, the constructed semantic graph of Python code contains less semantic information compared with that of Java code, which results in the worse performance. As such, increasing the semantic complexity of Python ASTs might be a practical way to achieve further performance improvements.

\subsubsection{Results on the Cleaned Testing Sets}
The above datasets are widely used in the current research for code summarization. However, we carefully examine the datasets and find that a large fraction of code snippets in the testing set are also presented in the training set, for both datasets. 

\begin{figure}[!htb]
  \centering
   \includegraphics[width=0.5\textwidth]{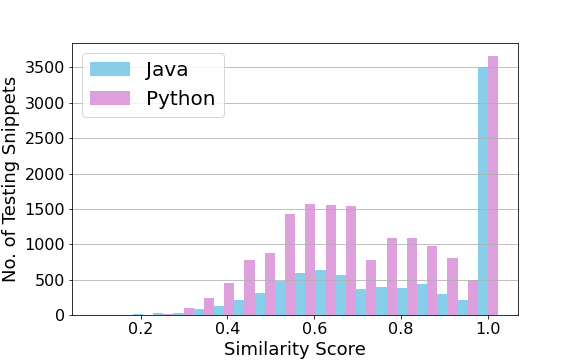}
  \caption{The distribution of similarity score on the two testing sets. The x-axis is the similarity score for each code snippet in the testing set. The y-axis is the count of code snippets.}
	\label{fig:overlap_ratio}
\end{figure}

In particular, for each code snippet in the testing set, we use the longest common sequence algorithm to find the most similar snippet in the training set and compute the similarity score between them. A score of $1.0$ indicates the two snippets are exactly the same. Then we plot the distribution of the similarity scores in Figure~\ref{fig:overlap_ratio} for all the testing code snippets in the two datasets. We observe that for the Java dataset, about $39\%$ of the code snippets in the testing set are actually contained in the training set (with similarity score equal to 1.0), and for the Python dataset, the fraction is about $21\%$.

To better evaluate the generalization ability of the models, we remove those duplicated snippets with similarity score 1 from the testing sets and report the additional results on the cleaned testing sets. We run those models that have released executable source code and the pre-trained models on the sets. The results are reported in Table~\ref{tab:remove}. We observe a drastic performance decrease for all the models. Nevertheless, GypSum still performs the best among all the models.
\begin{table*}[t]
  \centering
  \caption{The comparative results on the cleaned testing sets.}
  \label{tab:remove}
  \begin{tabular}{l|c|c|c|c|c|c}
    \hline
    \multirow{2}{*}{Methods}&\multicolumn{3}{c|}{Java}&\multicolumn{3}{c}{Python}\\
    \cline{2-7}
    &BLEU&METEOR&ROUGE-L&BLEU&METEOR&ROUGE-L\\
        \hline
    RL+HybridSeq (2018) \cite{wan2018improving}&14.99&11.01&28.99&11.12&7.21&30.31\\
    DeepCom (2018)\cite{hu2018deep}&15.20&11.19&29.03&12.03&7.29&29.98\\
    API+CODE (2018)\cite{hu2018summarizing}&15.93&11.56&28.94&11.39&6.99&23.68\\
    Dual (2019)\cite{wei2019code}&16.66&12.53&30.42&15.91&10.03&34.15\\
    Rencos (2020)\cite{zhang2020retrieval}&18.69&11.96&32.20&21.55&15.25&36.40\\
    Transformer (2020) \cite{ahmad-etal-2020-transformer}&18.46&13.90&31.76&22.74&14.26&38.52\\
    SiT (2021)\cite{wu2021code}&18.74&13.02&32.61&22.96&15.14&39.77\\
    CodeT5 (2021)
    \cite{wang2021codet5}&19.42&13.67&$\mathit{33.91}$&24.07&16.88&41.29\\
    CodeBERT (2020) \cite{feng2020codebert}&$\mathit{19.47}$&$\mathit{13.75}$&33.86&$\mathit{24.12}$&$\mathit{16.95}$&$\mathit{41.34}$\\
    \hline
    \multirow{2}{*}{\bf GypSum}&{\bf 21.35}&{\bf 15.11}&{\bf 37.32}&{\bf 25.22}&{\bf 17.59}&{\bf 42.10}\\
    &($\uparrow$ 1.88)&($\uparrow$ 1.36)&($\uparrow$ 3.41)&($\uparrow$ 1.10)&($\uparrow$ 0.64)&($\uparrow$ 0.76)\\
    \hline
    \end{tabular}
\end{table*}

\subsubsection{Ablation Study}\label{sec:ablation}
The main idea of GypSum is to use two encoders to learn simultaneously from token sequences and AST structures, with the expectation of preserving both naturalness and informativeness in the summaries. There can be different choices of the detailed structures for the encoders. As such we replace the two encoders with alternative structures and show GypSum performs the best among the possible structure choices. The results are presented in the bottom half of Table~\ref{tab:full}. 

For learning from the token sequences, we replace the c-encoder with the LSTM encoder and Transformer's encoder, respectively, and keep the g-encoder unchanged, which results in two models LSTM+g-encoder and Transformer+g-encoder. For learning from the AST structures, we keep the c-encoder unchanged and replace the g-encoder with the encoder in Code2Seq~\cite{alon2018codeseq}, ASTNN~\cite{zhang2019novel} and GGNN~\cite{allamanis2018learning}, respectively, which are three other state-of-the-art techniques for code representation learning using ASTs. Code2Seq extracts the paths from an AST and encodes each individual path. Hence we use the path encodings for the attention computation on the decoder side. ASTNN encodes the individual subtree in an AST corresponding to a set of statements in the source code. Similarly, we use the subtree encodings for the attention computation on the decoder side. GGNN constructs a similar semantic graph to ours for a code snippet and uses gated graph sequence neural networks to learn from the graphs. The resulted three models are c-encoder+Code2Seq, c-encoder+ASTNN and c-encoder+GGNN.

We observe two points. First, all the ablation models except LSTM+g-encoder outperform all the comparable methods. The results show again the effectiveness of learning hybrid representations from source code for summarization. Notably, Transformer+g-encoder also outperforms all the comparable models, indicating that with our method, we can achieve the new state of the art without relying on any pre-trained programming and natural language models, such as CodeBERT and CodeT5. LSTM+g-encoder performs worse than Transformer+g-encoder, since it has the worse ability to encode long sequences. Second, the proposed GypSum model outperforms all ablation models, which justifies the choice of each component among alternatives. GypSum is better than LSTM+g-encoder and Transformer+g-encoder, thanks to the ability of CodeBERT to produce more fluent natural language. For c-encoder+Code2Seq and c-encoder+ASTNN, they manipulate the AST structures without considering the direct links between the elements in the source code. For c-encoder+GGNN, it does not add the semantic edges proposed by us and has not considered to use attentions for node feature extraction. As such, the three ablation models perform worse than GypSum.

\subsubsection{Sensitivity Analysis}
\begin{table}
\centering
 \scriptsize
  \caption{Sensitivity analysis with different hyperparameter settings of GypSum.}
  \label{tab:parameter}
  \begin{tabular}{c|c|c|c|c|c|c|c}
  \hline
    \multicolumn{2}{c|}{}&\multicolumn{3}{c|}{Java}&\multicolumn{3}{c}{Python}\\
    \hline
    & &BLEU&METEOR&ROUGE-L&BLEU&METEOR&ROUGE-L\\
    \hline
    \multirow{4}{*}{$d_e$}  &192&39.66&22.83&52.28&28.13&17.58&42.89\\
    &384&43.39&26.74&55.87&30.42&19.77&45.12\\
    &576&46.21&28.62&57.36&33.69&21.41&48.88\\
    &768&\textbf{48.57}&\textbf{30.85}&\textbf{59.42}&\textbf{35.88}&\textbf{22.96}&\textbf{50.27}\\
    \hline
    \multicolumn{8}{c}{}\\
    \hline
    \multirow{4}{*}{$L_d$}&3&46.33&27.42&56.74&33.99&20.79&47.50\\
    &4&47.06&28.05&57.48&34.43&21.89&48.53\\
    &5&47.83&28.92&58.37&35.14&21.93&49.39\\
    &6&\textbf{48.57}&\textbf{30.85}&\textbf{59.42}&\textbf{35.88}&\textbf{22.96}&\textbf{50.27}\\
    \hline
    \multicolumn{8}{c}{}\\
    \hline  
    \multirow{6}{*}{$L_g$}&1&48.15&30.57&59.03&35.60&22.68&50.11\\
    &2&48.28&30.63&59.14&35.67&22.77&50.12\\
    &3&48.37&30.75&59.29&35.72&22.83&50.14\\
    &4&\textbf{48.57}&\textbf{30.85}&\textbf{59.42}&\textbf{35.88}&\textbf{22.96}&\textbf{50.27}\\
    &5&48.45&30.80&59.34&35.77&22.89&50.19\\
    &6&48.51&30.84&59.37&35.82&22.94&50.25\\
    \hline
    \multicolumn{8}{c}{}\\
    \hline     
     \multirow{4}{*}{$l_g$}&100&48.11&30.58&59.07&35.57&22.62&50.09\\
    &200&48.23&30.60&59.15&35.64&22.78&50.15\\
    &300&\textbf{48.57}&\textbf{30.85}&\textbf{59.42}&\textbf{35.88}&\textbf{22.96}&\textbf{50.27}\\  
    &400&48.55&30.81&59.38&35.85&22.92&50.23\\
    \hline
    \end{tabular}
\end{table}
We choose four most important hyperparameters in GypSum for sensitivity analysis, namely, the output embedding size of the two encoders $d_e$, the number of attention layers used in the decoder $L_d$, the graph attention layers in the g-encoder $L_g$, and the number of node embeddings output by the g-encoder $l_g$. $d_e$ is also the embedding size of the encoder-decoder attention sublayer in the decoder. Other hyperparameters such as the input length, the number of attention heads and the beam search size take the common settings in the related literature. The results are presented in Table~\ref{tab:parameter}. We vary $d_e$ in $\{192,384,576,768\}$. The results show that longer embedding size yields better performance, since more information of source code is preserved. We vary $L_d$ from 3 to 6 in increments of 1. The best performance is achieved at $L_d=6$, indicating that the deeper c-encoder can better capture the semantic information of the code. We vary $L_g$ from 1 to 6 and $l_g$ from 100 to 400 in increments of 1 and 100, respectively. We observe that the performance of GypSum is relatively stable when we vary $L_g$ and $l_g$. For $L_g$, the best performance is achieved at $L_g=4$, indicating that we do not have to use very deep layers for the graph attention neural networks. For $l_g$, the best performance is achieved at $l_g=300$, indicating that we do not have to use all the node information when learning from the AST-based graphs. This would reduce the training overhead greatly.

\subsection{User Study}
We conduct a user study on the quality of the summaries generated by the five best models in Table~\ref{tab:remove}, namely, Transformer, SiT, CodeT5, CodeBERT and our GypSum. The evaluated metrics are \textit{naturalness} and \textit{informativeness}. The former measures text quality pertaining to grammaticality and fluency, and the latter measures how much functional information is carried in a summary. We invite four PhD students and six Master's students whose major is computer science to rate the summaries. We randomly choose 100 code snippets from the testing sets (50 for Java and 50 for Python) and generate the summaries for each snippet using the five models. We replicate three times each snippet with its five summaries so that each student is assigned with 30 different snippets with 150 summaries. We ask the students to rate each summary on the two metrics on a scale between 1 to 3, where a higher score means better quality. Then we compute the average scores for each model and the statistical significance between GypSum and other models. The results are reported in Table~\ref{tab:human_eval}. We observe that for both datasets, GypSum performs the best on both metrics and the difference between GypSum and each model is statistically significant ($p\mbox{-}value<0.05$), verified using a 2-tailed Student’s t-test. Note that the initial goal of GypSum is to ensure both the naturalness and informativeness of the summary by learning hybrid representations of source code. The results justify the effectiveness of hybrid representation learning.

\subsection{Case Study}
\begin{table}
    \centering\small
    \begin{tabular}{c|c|c|c}
  \hline
    & \textbf{Model} & \textbf{Naturalness} & \textbf{Informativeness}\\
    \hline
    \parbox[t]{1mm}{\multirow{5}{*}{\rotatebox[origin=c]{90}{Java}}} & Transformer & 1.93 & 1.98\\
     & SiT & 2.06 & 2.11 \\
     & CodeT5 & 2.21 & 2.09 \\
     & CodeBERT & 2.38 & 2.22  \\
     & \bf GypSum  & \textbf{2.54} & \textbf{2.56} \\\hline
     \parbox[t]{1mm}{\multirow{5}{*}{\rotatebox[origin=c]{90}{Python}}} & Transformer  & 1.91 & 1.84  \\
     &  SiT  & 2.03 & 1.92 \\
     & CodeT5 & 2.30 & 1.97 \\
     & CodeBERT & 2.21 & 2.04\\
     & \bf GypSum  & \textbf{2.41} & \textbf{2.28} \\
    \hline
    \end{tabular}
    \caption{Naturalness and Informativeness Measurement.}
    \label{tab:human_eval}
\end{table}
\begin{figure}[!htb]
  \centering
  \subfigure[The heatmap of the first Java code from Figure~\ref{fig:sample_codes}.]{
\begin{minipage}[b]{0.43\textwidth}\label{fig:heat1}
\includegraphics[width=1\textwidth]{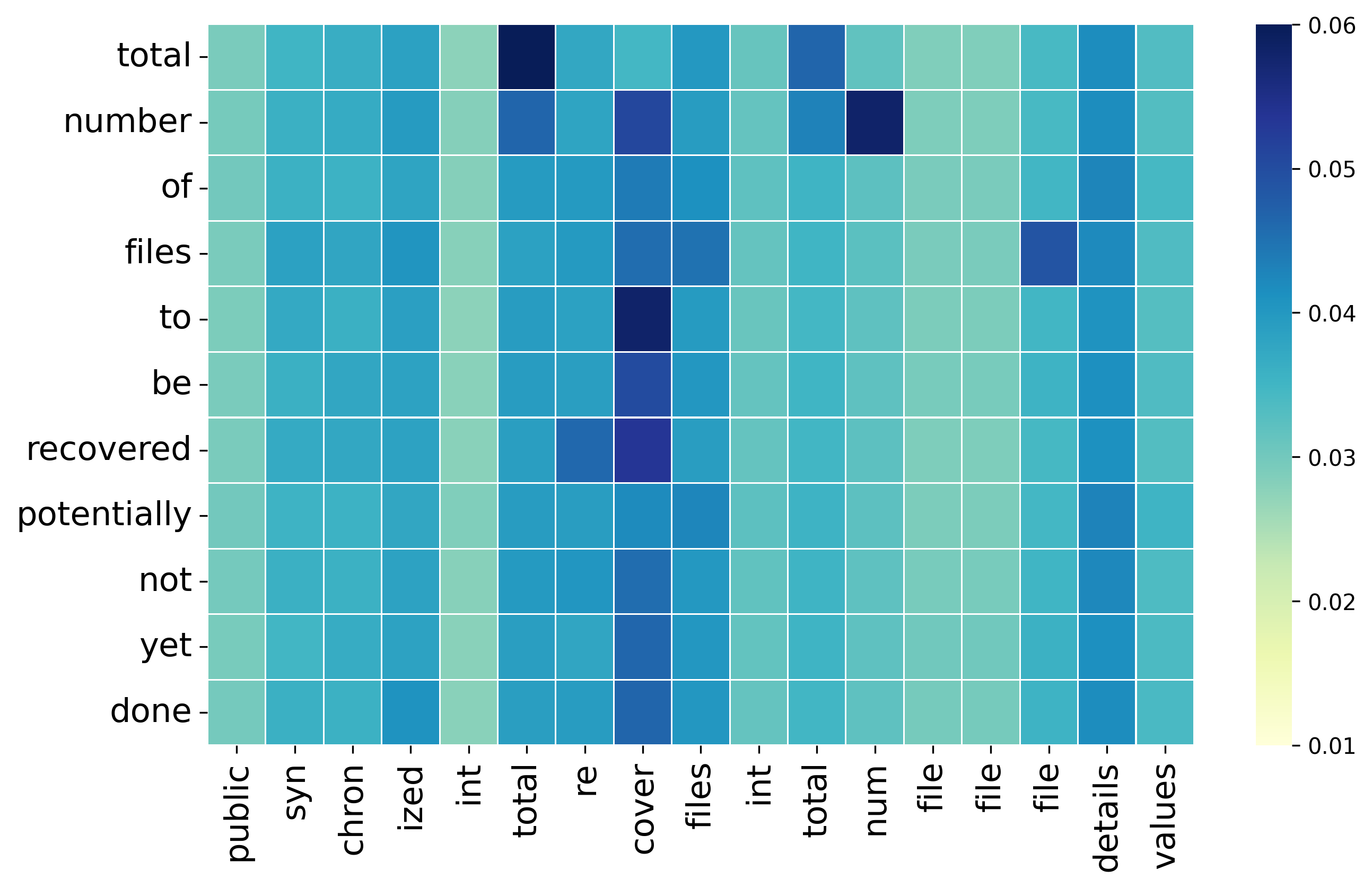}
\end{minipage}
}
\subfigure[The heatmap of the first Python code from Figure~\ref{fig:sample_codes}.]{
\begin{minipage}[b]{0.43\textwidth}\label{fig:heat2}
\includegraphics[width=1\textwidth]{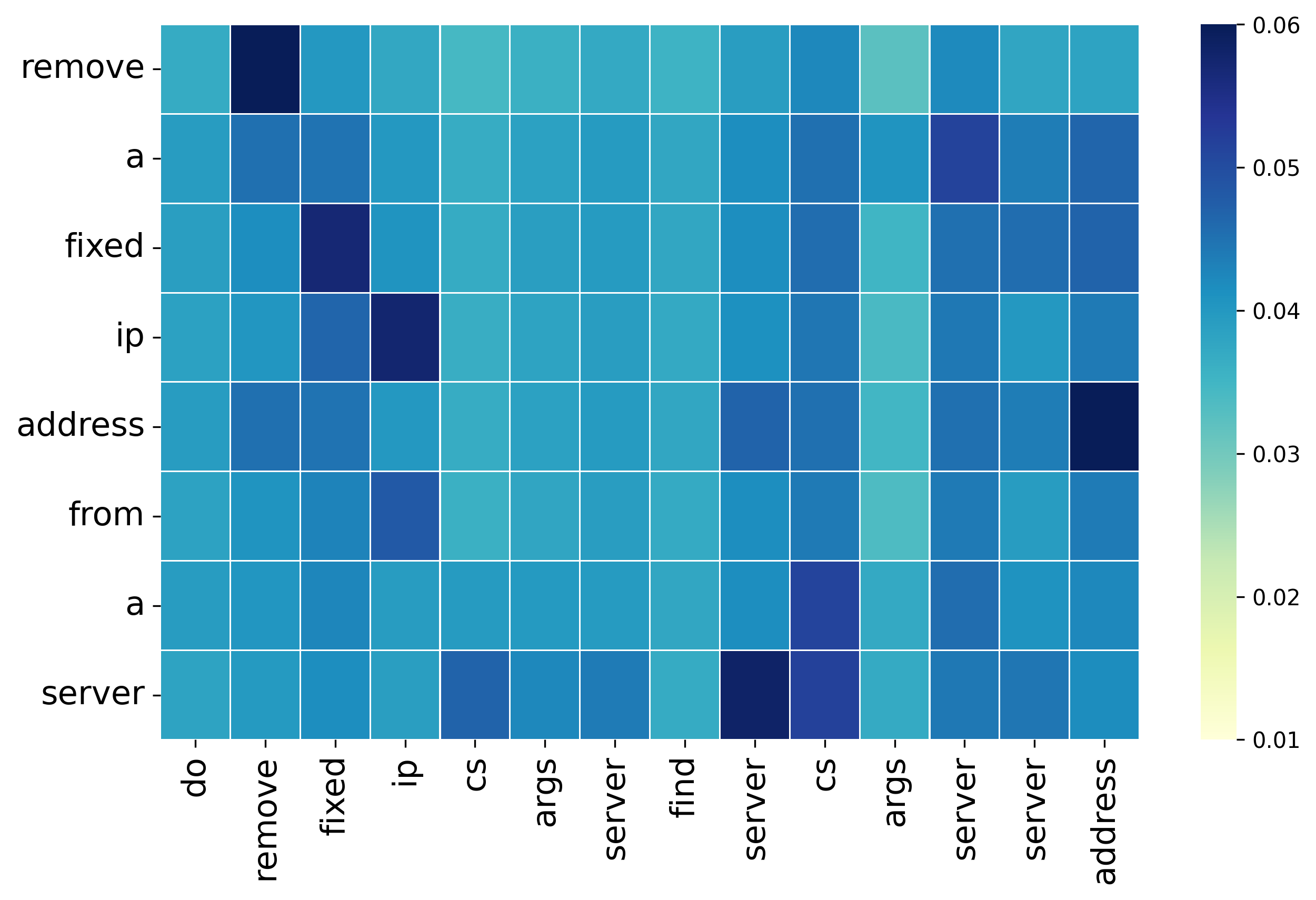}
\end{minipage}
}
  \caption{The total attention weight of each leaf node when producing each token in the corresponding summary. $X$-axis shows the leaf nodes in the AST and $Y$-axis is the summary. Each row shows how much every leaf node is attended when producing the token.}
	\label{fig:attentions}
\end{figure}
\subsubsection{Visualizing the Attention Weights}\label{sec:visual}
\begin{figure*}[h]
\centering
\includegraphics[width=1\textwidth]{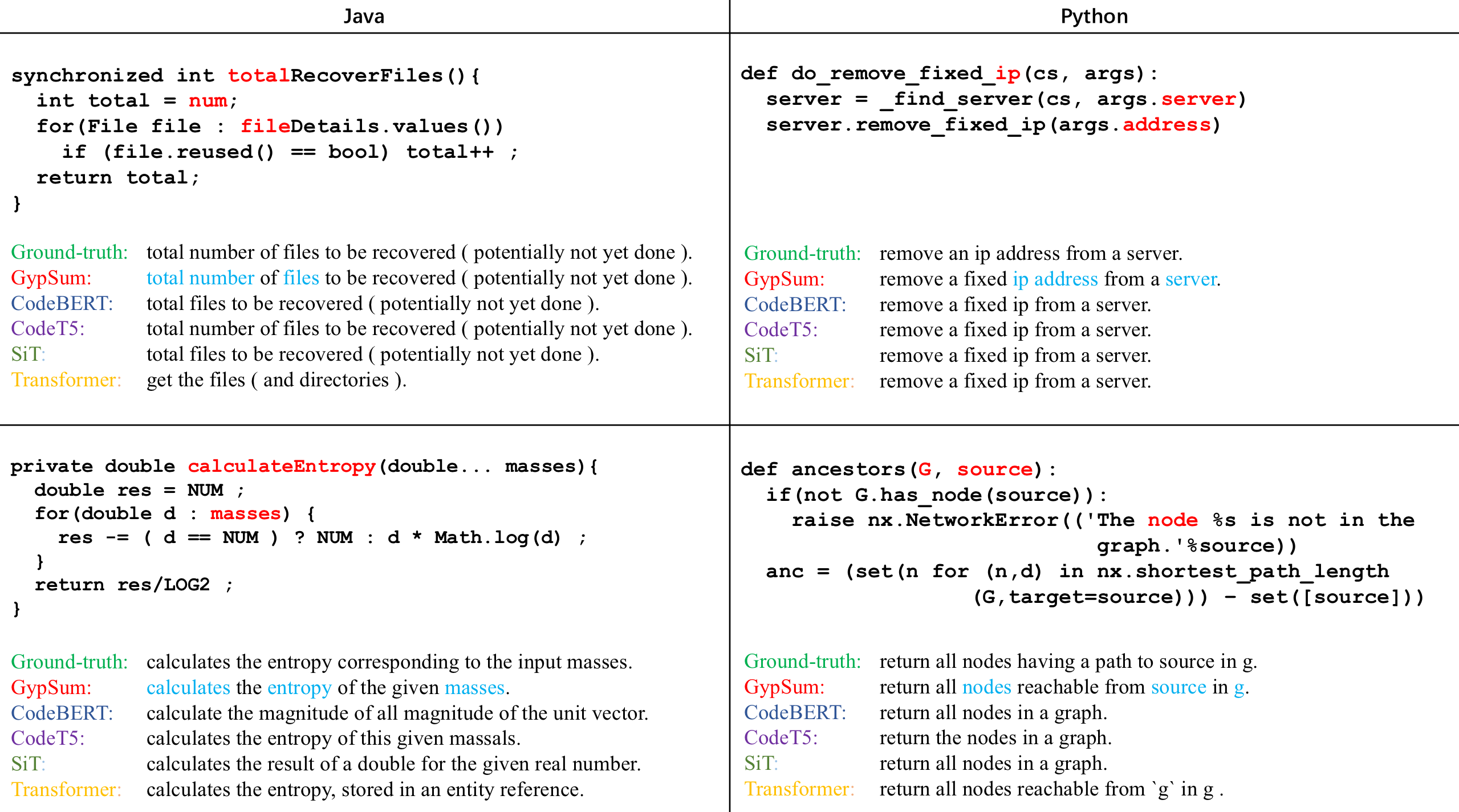}
\caption{Case study on four example code snippets. We mark the strings in red in the source code pertaining to the key functionality of each code snippet, and mark the words in light-blue in the summaries generated by GypSum that capture the key functionality of the corresponding code snippet.}
\label{fig:sample_codes}
\end{figure*}

Since our motivation is to improve the model ability for generating tokens that capture the key functionality of a code snippet, we investigate what elements are captured by the g-encoder when GypSum generates each token in the summary. To this end, we calculate the total attention weight of each leaf node in the AST that contributes to the generation of each token in the summary. Particularly, the total attention weight of a leaf node $i$ contributing to a token $j$ is calculated as the sum of the multiplication between the attention weight of $i$ to each encoding of g-encoder and the attention weight of the encoding to the output token $j$ in the decoder. We calculate the attention weights in the last layer of the model. We visualize the attention weights of the top two code snippets in Figure~\ref{fig:sample_codes}, and present the result in Figure~\ref{fig:heat1} and Figure~\ref{fig:heat2}. The $X$-axis is the (partial) code snippet and the $Y$-axis is the summary. A deeper color indicates a higher total attention weight. We observe that when producing some key tokens, the model successfully attends to the corresponding leaf nodes through GAT. For instance, in Figure~\ref{fig:heat1}, the model particularly attends to the leaf node `total' and `num' when generating `total' and `number' in the summary, respectively.

\subsubsection{Code Summary Examples}
We present some summary examples generated by different models in Figure~\ref{fig:sample_codes}, including GypSum, CodeBERT, CodeT5, SiT and Transformer. The Transformer model learns only from the token sequence of a code snippet. As expected, we observe that Transformer can generate summaries of good grammar structure. However, the summaries do not capture well the functional information in the code snippets. For example at the top-left corner, the summary generated by Transformer is `\emph{get the files (and directories)}' whereas the true functionality is to count the total number of the recovered files. For the two examples at the bottom, the Transformer model generates meaningless summaries. SiT improves Transformer by introducing node relationship in the constructed graph into the attention computation. However, the example in the bottom-left corner shows it can still miss the key functional information and generate an incorrect summary. On the other hand, the two pre-trained models benefit from the pre-training tasks using code snippets and the summaries, and produce much better results. For example at the top-left corner, CodeT5 could produce exactly the same summary as the ground-truth. On all the four examples, GypSum generates the summaries of the overall best quality. It not only captures the key functional information (highlighted using red color in the code snippets and using light-blue color in the summaries), but also produces very fluent summaries, which are highly close to the ground-truth text.

\section{Conclusion and Future Work}
\label{sec:conclusion}
In this work, we show that learning hybrid representations of source code can produce better summaries of the code. We follow the encoder-decoder architecture and propose GypSum, a new deep learning model for the task. GypSum has two encoders, one leveraging the graph attention neural networks to learn from semantic graphs constructed using ASTs, and the other adopting a recently developed pre-trained PL-NL model to learn from token sequences. We use the Transformer's decoder to generate summaries, where we modify the encoder-decoder sublayer to fuse the output of the two encoders. We conduct extensive experiments to show the superior performance of GypSum over existing models and justify the choice of each component in GypSum.

Given there are other representation learning methods of source code, in future, we will investigate the effect of more code representations and other methods for hybrid representation learning. One possible direction is to employ automated machine learning methods to identify the optimal code representations for code summarization as well as the architecture for representation fusion.

\section*{Acknowledgement}
This work is supported by the grants from the National Natural Science Foundation of China (Grant No. 61977026, 62072185).

\nocite{iyer2016summarizing}
\nocite{eriguchi2016tree}
\nocite{wan2018improving}
\nocite{hu2018deep}
\nocite{hu2018summarizing}
\nocite{ahmad-etal-2020-transformer}
\nocite{alon2018codeseq}
\nocite{vaswani2017attention}
\nocite{papineni2002bleu}
\nocite{banerjee2005meteor}
\nocite{lin-2004-rouge}
\nocite{kingma2015adam}
\nocite{leclair2020improved}
\nocite{shido2019automatic}
\nocite{devlin2019bert}
\nocite{radfordimproving}
\nocite{hindle2012naturalness}
\nocite{radfordimproving}
\nocite{allamanis2015suggesting}
\nocite{feng2020codebert}
\nocite{wei2019code}



\clearpage
\bibliographystyle{ACM-Reference-Format}
\bibliography{reference}


\begin{thebibliography}{48}


\ifx \showCODEN    \undefined \def \showCODEN     #1{\unskip}     \fi
\ifx \showDOI      \undefined \def \showDOI       #1{#1}\fi
\ifx \showISBNx    \undefined \def \showISBNx     #1{\unskip}     \fi
\ifx \showISBNxiii \undefined \def \showISBNxiii  #1{\unskip}     \fi
\ifx \showISSN     \undefined \def \showISSN      #1{\unskip}     \fi
\ifx \showLCCN     \undefined \def \showLCCN      #1{\unskip}     \fi
\ifx \shownote     \undefined \def \shownote      #1{#1}          \fi
\ifx \showarticletitle \undefined \def \showarticletitle #1{#1}   \fi
\ifx \showURL      \undefined \def \showURL       {\relax}        \fi
\providecommand\bibfield[2]{#2}
\providecommand\bibinfo[2]{#2}
\providecommand\natexlab[1]{#1}
\providecommand\showeprint[2][]{arXiv:#2}

\bibitem[Ahmad et~al\mbox{.}(2020)]%
        {ahmad-etal-2020-transformer}
\bibfield{author}{\bibinfo{person}{Wasi Ahmad}, \bibinfo{person}{Saikat
  Chakraborty}, \bibinfo{person}{Baishakhi Ray}, {and} \bibinfo{person}{Kai-Wei
  Chang}.} \bibinfo{year}{2020}\natexlab{}.
\newblock \showarticletitle{A Transformer-based Approach for Source Code
  Summarization}. In \bibinfo{booktitle}{\emph{Proceedings of the 58th Annual
  Meeting of the Association for Computational Linguistics}}.
  \bibinfo{pages}{4998--5007}.
\newblock


\bibitem[Allamanis et~al\mbox{.}(2015)]%
        {allamanis2015suggesting}
\bibfield{author}{\bibinfo{person}{Miltiadis Allamanis},
  \bibinfo{person}{Earl~T Barr}, \bibinfo{person}{Christian Bird}, {and}
  \bibinfo{person}{Charles Sutton}.} \bibinfo{year}{2015}\natexlab{}.
\newblock \showarticletitle{Suggesting accurate method and class names}. In
  \bibinfo{booktitle}{\emph{Proceedings of the 2015 10th Joint Meeting on
  Foundations of Software Engineering}}. \bibinfo{pages}{38--49}.
\newblock


\bibitem[Allamanis et~al\mbox{.}(2018)]%
        {allamanis2018learning}
\bibfield{author}{\bibinfo{person}{Miltiadis Allamanis}, \bibinfo{person}{Marc
  Brockschmidt}, {and} \bibinfo{person}{Mahmoud Khademi}.}
  \bibinfo{year}{2018}\natexlab{}.
\newblock \showarticletitle{Learning to Represent Programs with Graphs}. In
  \bibinfo{booktitle}{\emph{International Conference on Learning
  Representations}}.
\newblock


\bibitem[Alon et~al\mbox{.}(2019a)]%
        {alon2018codeseq}
\bibfield{author}{\bibinfo{person}{Uri Alon}, \bibinfo{person}{Omer Levy},
  {and} \bibinfo{person}{Eran Yahav}.} \bibinfo{year}{2019}\natexlab{a}.
\newblock \showarticletitle{code2seq: Generating Sequences from Structured
  Representations of Code}. In \bibinfo{booktitle}{\emph{International
  Conference on Learning Representations}}.
\newblock


\bibitem[Alon et~al\mbox{.}(2018)]%
        {alon2018general}
\bibfield{author}{\bibinfo{person}{Uri Alon}, \bibinfo{person}{Meital
  Zilberstein}, \bibinfo{person}{Omer Levy}, {and} \bibinfo{person}{Eran
  Yahav}.} \bibinfo{year}{2018}\natexlab{}.
\newblock \showarticletitle{A general path-based representation for predicting
  program properties}.
\newblock \bibinfo{journal}{\emph{ACM SIGPLAN Notices}} \bibinfo{volume}{53},
  \bibinfo{number}{4} (\bibinfo{year}{2018}), \bibinfo{pages}{404--419}.
\newblock


\bibitem[Alon et~al\mbox{.}(2019b)]%
        {alon2019code2vec}
\bibfield{author}{\bibinfo{person}{Uri Alon}, \bibinfo{person}{Meital
  Zilberstein}, \bibinfo{person}{Omer Levy}, {and} \bibinfo{person}{Eran
  Yahav}.} \bibinfo{year}{2019}\natexlab{b}.
\newblock \showarticletitle{code2vec: Learning distributed representations of
  code}.
\newblock \bibinfo{journal}{\emph{Proceedings of the ACM on Programming
  Languages}} \bibinfo{volume}{3}, \bibinfo{number}{POPL}
  (\bibinfo{year}{2019}), \bibinfo{pages}{1--29}.
\newblock


\bibitem[Bahdanau et~al\mbox{.}(2015)]%
        {bahdanau2015neural}
\bibfield{author}{\bibinfo{person}{Dzmitry Bahdanau},
  \bibinfo{person}{Kyung~Hyun Cho}, {and} \bibinfo{person}{Yoshua Bengio}.}
  \bibinfo{year}{2015}\natexlab{}.
\newblock \showarticletitle{Neural machine translation by jointly learning to
  align and translate}. In \bibinfo{booktitle}{\emph{3rd International
  Conference on Learning Representations, ICLR 2015}}.
\newblock


\bibitem[Banerjee and Lavie(2005)]%
        {banerjee2005meteor}
\bibfield{author}{\bibinfo{person}{Satanjeev Banerjee} {and}
  \bibinfo{person}{Alon Lavie}.} \bibinfo{year}{2005}\natexlab{}.
\newblock \showarticletitle{METEOR: An automatic metric for MT evaluation with
  improved correlation with human judgments}. In
  \bibinfo{booktitle}{\emph{Proceedings of the acl workshop on intrinsic and
  extrinsic evaluation measures for machine translation and/or summarization}}.
  \bibinfo{pages}{65--72}.
\newblock


\bibitem[Brockschmidt et~al\mbox{.}(2018)]%
        {brockschmidt2018generative}
\bibfield{author}{\bibinfo{person}{Marc Brockschmidt},
  \bibinfo{person}{Miltiadis Allamanis}, \bibinfo{person}{Alexander~L Gaunt},
  {and} \bibinfo{person}{Oleksandr Polozov}.} \bibinfo{year}{2018}\natexlab{}.
\newblock \showarticletitle{Generative code modeling with graphs}.
\newblock \bibinfo{journal}{\emph{arXiv preprint arXiv:1805.08490}}
  (\bibinfo{year}{2018}).
\newblock


\bibitem[Chen et~al\mbox{.}(2021)]%
        {chen2021novel}
\bibfield{author}{\bibinfo{person}{Fuxiang Chen}, \bibinfo{person}{Mijung Kim},
  {and} \bibinfo{person}{Jaegul Choo}.} \bibinfo{year}{2021}\natexlab{}.
\newblock \showarticletitle{Novel Natural Language Summarization of Program
  Code via Leveraging Multiple Input Representations}. In
  \bibinfo{booktitle}{\emph{Findings of the Association for Computational
  Linguistics: EMNLP 2021}}. \bibinfo{pages}{2510--2520}.
\newblock


\bibitem[Choi et~al\mbox{.}(2021)]%
        {choi2021learning}
\bibfield{author}{\bibinfo{person}{YunSeok Choi}, \bibinfo{person}{JinYeong
  Bak}, \bibinfo{person}{CheolWon Na}, {and} \bibinfo{person}{Jee-Hyong Lee}.}
  \bibinfo{year}{2021}\natexlab{}.
\newblock \showarticletitle{Learning Sequential and Structural Information for
  Source Code Summarization}. In \bibinfo{booktitle}{\emph{Findings of the
  Association for Computational Linguistics: ACL-IJCNLP 2021}}.
  \bibinfo{pages}{2842--2851}.
\newblock


\bibitem[Clark et~al\mbox{.}(2019)]%
        {clark2019electra}
\bibfield{author}{\bibinfo{person}{Kevin Clark}, \bibinfo{person}{Minh-Thang
  Luong}, \bibinfo{person}{Quoc~V Le}, {and} \bibinfo{person}{Christopher~D
  Manning}.} \bibinfo{year}{2019}\natexlab{}.
\newblock \showarticletitle{ELECTRA: Pre-training Text Encoders as
  Discriminators Rather Than Generators}. In
  \bibinfo{booktitle}{\emph{International Conference on Learning
  Representations}}.
\newblock


\bibitem[Cvitkovic et~al\mbox{.}(2019)]%
        {cvitkovic2019open}
\bibfield{author}{\bibinfo{person}{Milan Cvitkovic}, \bibinfo{person}{Badal
  Singh}, {and} \bibinfo{person}{Animashree Anandkumar}.}
  \bibinfo{year}{2019}\natexlab{}.
\newblock \showarticletitle{Open vocabulary learning on source code with a
  graph-structured cache}. In \bibinfo{booktitle}{\emph{International
  Conference on Machine Learning}}. PMLR, \bibinfo{pages}{1475--1485}.
\newblock


\bibitem[de~Souza et~al\mbox{.}(2005)]%
        {de2005study}
\bibfield{author}{\bibinfo{person}{Sergio Cozzetti~B de Souza},
  \bibinfo{person}{Nicolas Anquetil}, {and} \bibinfo{person}{K{\'a}thia~M de
  Oliveira}.} \bibinfo{year}{2005}\natexlab{}.
\newblock \showarticletitle{A study of the documentation essential to software
  maintenance}. In \bibinfo{booktitle}{\emph{Proceedings of the 23rd annual
  international conference on Design of communication: documenting \& designing
  for pervasive information}}. \bibinfo{pages}{68--75}.
\newblock


\bibitem[Devlin et~al\mbox{.}(2019)]%
        {devlin2019bert}
\bibfield{author}{\bibinfo{person}{Jacob Devlin}, \bibinfo{person}{Ming-Wei
  Chang}, \bibinfo{person}{Kenton Lee}, {and} \bibinfo{person}{Kristina
  Toutanova}.} \bibinfo{year}{2019}\natexlab{}.
\newblock \showarticletitle{BERT: Pre-training of Deep Bidirectional
  Transformers for Language Understanding}. In
  \bibinfo{booktitle}{\emph{Proceedings of the 2019 Conference of the North
  American Chapter of the Association for Computational Linguistics: Human
  Language Technologies, Volume 1 (Long and Short Papers)}}.
  \bibinfo{pages}{4171--4186}.
\newblock


\bibitem[Eriguchi et~al\mbox{.}(2016)]%
        {eriguchi2016tree}
\bibfield{author}{\bibinfo{person}{Akiko Eriguchi}, \bibinfo{person}{Kazuma
  Hashimoto}, {and} \bibinfo{person}{Yoshimasa Tsuruoka}.}
  \bibinfo{year}{2016}\natexlab{}.
\newblock \showarticletitle{Tree-to-Sequence Attentional Neural Machine
  Translation}. In \bibinfo{booktitle}{\emph{Proceedings of the 54th Annual
  Meeting of the Association for Computational Linguistics (Volume 1: Long
  Papers)}}. \bibinfo{pages}{823--833}.
\newblock


\bibitem[Feng et~al\mbox{.}(2020)]%
        {feng2020codebert}
\bibfield{author}{\bibinfo{person}{Zhangyin Feng}, \bibinfo{person}{Daya Guo},
  \bibinfo{person}{Duyu Tang}, \bibinfo{person}{Nan Duan},
  \bibinfo{person}{Xiaocheng Feng}, \bibinfo{person}{Ming Gong},
  \bibinfo{person}{Linjun Shou}, \bibinfo{person}{Bing Qin},
  \bibinfo{person}{Ting Liu}, \bibinfo{person}{Daxin Jiang}, {et~al\mbox{.}}}
  \bibinfo{year}{2020}\natexlab{}.
\newblock \showarticletitle{CodeBERT: A Pre-Trained Model for Programming and
  Natural Languages}. In \bibinfo{booktitle}{\emph{Proceedings of the 2020
  Conference on Empirical Methods in Natural Language Processing: Findings}}.
  \bibinfo{pages}{1536--1547}.
\newblock


\bibitem[Gao et~al\mbox{.}(2021)]%
        {gao2021code}
\bibfield{author}{\bibinfo{person}{Shuzheng Gao}, \bibinfo{person}{Cuiyun Gao},
  \bibinfo{person}{Yulan He}, \bibinfo{person}{Jichuan Zeng},
  \bibinfo{person}{Lun~Yiu Nie}, {and} \bibinfo{person}{Xin Xia}.}
  \bibinfo{year}{2021}\natexlab{}.
\newblock \showarticletitle{Code Structure Guided Transformer for Source Code
  Summarization}.
\newblock \bibinfo{journal}{\emph{arXiv preprint arXiv:2104.09340}}
  (\bibinfo{year}{2021}).
\newblock


\bibitem[Haque et~al\mbox{.}(2020)]%
        {haque2020improved}
\bibfield{author}{\bibinfo{person}{Sakib Haque}, \bibinfo{person}{Alexander
  LeClair}, \bibinfo{person}{Lingfei Wu}, {and} \bibinfo{person}{Collin
  McMillan}.} \bibinfo{year}{2020}\natexlab{}.
\newblock \showarticletitle{Improved automatic summarization of subroutines via
  attention to file context}. In \bibinfo{booktitle}{\emph{Proceedings of the
  17th International Conference on Mining Software Repositories}}.
  \bibinfo{pages}{300--310}.
\newblock


\bibitem[Hindle et~al\mbox{.}(2012)]%
        {hindle2012naturalness}
\bibfield{author}{\bibinfo{person}{Abram Hindle}, \bibinfo{person}{Earl~T
  Barr}, \bibinfo{person}{Zhendong Su}, \bibinfo{person}{Mark Gabel}, {and}
  \bibinfo{person}{Premkumar Devanbu}.} \bibinfo{year}{2012}\natexlab{}.
\newblock \showarticletitle{On the naturalness of software}. In
  \bibinfo{booktitle}{\emph{2012 34th International Conference on Software
  Engineering (ICSE)}}. IEEE, \bibinfo{pages}{837--847}.
\newblock


\bibitem[Hu et~al\mbox{.}(2018a)]%
        {hu2018deep}
\bibfield{author}{\bibinfo{person}{Xing Hu}, \bibinfo{person}{Ge Li},
  \bibinfo{person}{Xin Xia}, \bibinfo{person}{David Lo}, {and}
  \bibinfo{person}{Zhi Jin}.} \bibinfo{year}{2018}\natexlab{a}.
\newblock \showarticletitle{Deep code comment generation}. In
  \bibinfo{booktitle}{\emph{2018 IEEE/ACM 26th International Conference on
  Program Comprehension (ICPC)}}. IEEE, \bibinfo{pages}{200--20010}.
\newblock


\bibitem[Hu et~al\mbox{.}(2018b)]%
        {hu2018summarizing}
\bibfield{author}{\bibinfo{person}{Xing Hu}, \bibinfo{person}{Ge Li},
  \bibinfo{person}{Xin Xia}, \bibinfo{person}{David Lo}, \bibinfo{person}{Shuai
  Lu}, {and} \bibinfo{person}{Zhi Jin}.} \bibinfo{year}{2018}\natexlab{b}.
\newblock \showarticletitle{Summarizing source code with transferred API
  knowledge}. In \bibinfo{booktitle}{\emph{Proceedings of the 27th
  International Joint Conference on Artificial Intelligence}}.
  \bibinfo{pages}{2269--2275}.
\newblock


\bibitem[Iyer et~al\mbox{.}(2016)]%
        {iyer2016summarizing}
\bibfield{author}{\bibinfo{person}{Srinivasan Iyer}, \bibinfo{person}{Ioannis
  Konstas}, \bibinfo{person}{Alvin Cheung}, {and} \bibinfo{person}{Luke
  Zettlemoyer}.} \bibinfo{year}{2016}\natexlab{}.
\newblock \showarticletitle{Summarizing source code using a neural attention
  model}. In \bibinfo{booktitle}{\emph{Proceedings of the 54th Annual Meeting
  of the Association for Computational Linguistics (Volume 1: Long Papers)}}.
  \bibinfo{pages}{2073--2083}.
\newblock


\bibitem[Kingma and Ba(2015)]%
        {kingma2015adam}
\bibfield{author}{\bibinfo{person}{Diederik~P Kingma} {and}
  \bibinfo{person}{Jimmy Ba}.} \bibinfo{year}{2015}\natexlab{}.
\newblock \showarticletitle{Adam: A method for stochastic optimization}.
\newblock \bibinfo{journal}{\emph{International Conference on Learning
  Representations}} (\bibinfo{year}{2015}).
\newblock


\bibitem[LeClair et~al\mbox{.}(2020)]%
        {leclair2020improved}
\bibfield{author}{\bibinfo{person}{Alexander LeClair}, \bibinfo{person}{Sakib
  Haque}, \bibinfo{person}{Lingfei Wu}, {and} \bibinfo{person}{Collin
  McMillan}.} \bibinfo{year}{2020}\natexlab{}.
\newblock \showarticletitle{Improved code summarization via a graph neural
  network}. In \bibinfo{booktitle}{\emph{Proceedings of the 28th International
  Conference on Program Comprehension}}. \bibinfo{pages}{184--195}.
\newblock


\bibitem[LeClair et~al\mbox{.}(2019)]%
        {leclair2019neural}
\bibfield{author}{\bibinfo{person}{Alexander LeClair}, \bibinfo{person}{Siyuan
  Jiang}, {and} \bibinfo{person}{Collin McMillan}.}
  \bibinfo{year}{2019}\natexlab{}.
\newblock \showarticletitle{A neural model for generating natural language
  summaries of program subroutines}. In \bibinfo{booktitle}{\emph{2019 IEEE/ACM
  41st International Conference on Software Engineering (ICSE)}}. IEEE,
  \bibinfo{pages}{795--806}.
\newblock


\bibitem[Li et~al\mbox{.}(2015)]%
        {li2015gated}
\bibfield{author}{\bibinfo{person}{Yujia Li}, \bibinfo{person}{Daniel Tarlow},
  \bibinfo{person}{Marc Brockschmidt}, {and} \bibinfo{person}{Richard Zemel}.}
  \bibinfo{year}{2015}\natexlab{}.
\newblock \showarticletitle{Gated graph sequence neural networks}.
\newblock \bibinfo{journal}{\emph{arXiv preprint arXiv:1511.05493}}
  (\bibinfo{year}{2015}).
\newblock


\bibitem[Lin(2004)]%
        {lin-2004-rouge}
\bibfield{author}{\bibinfo{person}{Chin-Yew Lin}.}
  \bibinfo{year}{2004}\natexlab{}.
\newblock \showarticletitle{{ROUGE}: A Package for Automatic Evaluation of
  Summaries}. In \bibinfo{booktitle}{\emph{Text Summarization Branches Out}}.
  \bibinfo{publisher}{Association for Computational Linguistics},
  \bibinfo{address}{Barcelona, Spain}, \bibinfo{pages}{74--81}.
\newblock


\bibitem[Liu et~al\mbox{.}(2019)]%
        {liu2019roberta}
\bibfield{author}{\bibinfo{person}{Yinhan Liu}, \bibinfo{person}{Myle Ott},
  \bibinfo{person}{Naman Goyal}, \bibinfo{person}{Jingfei Du},
  \bibinfo{person}{Mandar Joshi}, \bibinfo{person}{Danqi Chen},
  \bibinfo{person}{Omer Levy}, \bibinfo{person}{Mike Lewis},
  \bibinfo{person}{Luke Zettlemoyer}, {and} \bibinfo{person}{Veselin
  Stoyanov}.} \bibinfo{year}{2019}\natexlab{}.
\newblock \showarticletitle{Roberta: A robustly optimized bert pretraining
  approach}.
\newblock \bibinfo{journal}{\emph{arXiv preprint arXiv:1907.11692}}
  (\bibinfo{year}{2019}).
\newblock


\bibitem[Papineni et~al\mbox{.}(2002)]%
        {papineni2002bleu}
\bibfield{author}{\bibinfo{person}{Kishore Papineni}, \bibinfo{person}{Salim
  Roukos}, \bibinfo{person}{Todd Ward}, {and} \bibinfo{person}{Wei-Jing Zhu}.}
  \bibinfo{year}{2002}\natexlab{}.
\newblock \showarticletitle{BLEU: a method for automatic evaluation of machine
  translation}. In \bibinfo{booktitle}{\emph{Proceedings of the 40th annual
  meeting of the Association for Computational Linguistics}}.
  \bibinfo{pages}{311--318}.
\newblock


\bibitem[Radford et~al\mbox{.}(2018)]%
        {radfordimproving}
\bibfield{author}{\bibinfo{person}{Alec Radford}, \bibinfo{person}{Karthik
  Narasimhan}, \bibinfo{person}{Tim Salimans}, {and} \bibinfo{person}{Ilya
  Sutskever}.} \bibinfo{year}{2018}\natexlab{}.
\newblock \showarticletitle{Improving Language Understanding by Generative
  Pre-Training}.
\newblock  (\bibinfo{year}{2018}).
\newblock


\bibitem[Raffel et~al\mbox{.}(2020)]%
        {raffel2020exploring}
\bibfield{author}{\bibinfo{person}{Colin Raffel}, \bibinfo{person}{Noam
  Shazeer}, \bibinfo{person}{Adam Roberts}, \bibinfo{person}{Katherine Lee},
  \bibinfo{person}{Sharan Narang}, \bibinfo{person}{Michael Matena},
  \bibinfo{person}{Yanqi Zhou}, \bibinfo{person}{Wei Li}, {and}
  \bibinfo{person}{Peter~J Liu}.} \bibinfo{year}{2020}\natexlab{}.
\newblock \showarticletitle{Exploring the Limits of Transfer Learning with a
  Unified Text-to-Text Transformer}.
\newblock \bibinfo{journal}{\emph{Journal of Machine Learning Research}}
  \bibinfo{volume}{21} (\bibinfo{year}{2020}), \bibinfo{pages}{1--67}.
\newblock


\bibitem[Raychev et~al\mbox{.}(2015)]%
        {raychev2015predicting}
\bibfield{author}{\bibinfo{person}{Veselin Raychev}, \bibinfo{person}{Martin
  Vechev}, {and} \bibinfo{person}{Andreas Krause}.}
  \bibinfo{year}{2015}\natexlab{}.
\newblock \showarticletitle{Predicting program properties from" big code"}.
\newblock \bibinfo{journal}{\emph{ACM SIGPLAN Notices}} \bibinfo{volume}{50},
  \bibinfo{number}{1} (\bibinfo{year}{2015}), \bibinfo{pages}{111--124}.
\newblock


\bibitem[Schrouff et~al\mbox{.}(2019)]%
        {schrouff2019inferring}
\bibfield{author}{\bibinfo{person}{Jessica Schrouff}, \bibinfo{person}{Kai
  Wohlfahrt}, \bibinfo{person}{Bruno Marnette}, {and} \bibinfo{person}{Liam
  Atkinson}.} \bibinfo{year}{2019}\natexlab{}.
\newblock \showarticletitle{Inferring javascript types using graph neural
  networks}.
\newblock \bibinfo{journal}{\emph{arXiv preprint arXiv:1905.06707}}
  (\bibinfo{year}{2019}).
\newblock


\bibitem[Shi et~al\mbox{.}(2021)]%
        {shi2021cast}
\bibfield{author}{\bibinfo{person}{Ensheng Shi}, \bibinfo{person}{Yanlin Wang},
  \bibinfo{person}{Lun Du}, \bibinfo{person}{Hongyu Zhang},
  \bibinfo{person}{Shi Han}, \bibinfo{person}{Dongmei Zhang}, {and}
  \bibinfo{person}{Hongbin Sun}.} \bibinfo{year}{2021}\natexlab{}.
\newblock \showarticletitle{CAST: Enhancing Code Summarization with
  Hierarchical Splitting and Reconstruction of Abstract Syntax Trees}. In
  \bibinfo{booktitle}{\emph{Proceedings of the 2021 Conference on Empirical
  Methods in Natural Language Processing}}. \bibinfo{pages}{4053--4062}.
\newblock


\bibitem[Shido et~al\mbox{.}(2019)]%
        {shido2019automatic}
\bibfield{author}{\bibinfo{person}{Yusuke Shido}, \bibinfo{person}{Yasuaki
  Kobayashi}, \bibinfo{person}{Akihiro Yamamoto}, \bibinfo{person}{Atsushi
  Miyamoto}, {and} \bibinfo{person}{Tadayuki Matsumura}.}
  \bibinfo{year}{2019}\natexlab{}.
\newblock \showarticletitle{Automatic source code summarization with extended
  tree-lstm}. In \bibinfo{booktitle}{\emph{2019 International Joint Conference
  on Neural Networks (IJCNN)}}. IEEE, \bibinfo{pages}{1--8}.
\newblock


\bibitem[Sutskever et~al\mbox{.}(2014)]%
        {sutskever2014sequence}
\bibfield{author}{\bibinfo{person}{Ilya Sutskever}, \bibinfo{person}{Oriol
  Vinyals}, {and} \bibinfo{person}{Quoc~V Le}.}
  \bibinfo{year}{2014}\natexlab{}.
\newblock \showarticletitle{Sequence to Sequence Learning with Neural
  Networks}.
\newblock \bibinfo{journal}{\emph{Advances in Neural Information Processing
  Systems}}  \bibinfo{volume}{27} (\bibinfo{year}{2014}),
  \bibinfo{pages}{3104--3112}.
\newblock


\bibitem[Vaswani et~al\mbox{.}(2017)]%
        {vaswani2017attention}
\bibfield{author}{\bibinfo{person}{Ashish Vaswani}, \bibinfo{person}{Noam
  Shazeer}, \bibinfo{person}{Niki Parmar}, \bibinfo{person}{Jakob Uszkoreit},
  \bibinfo{person}{Llion Jones}, \bibinfo{person}{Aidan~N Gomez},
  \bibinfo{person}{{\L}ukasz Kaiser}, {and} \bibinfo{person}{Illia
  Polosukhin}.} \bibinfo{year}{2017}\natexlab{}.
\newblock \showarticletitle{Attention is all you need}.
\newblock \bibinfo{journal}{\emph{Advances in neural information processing
  systems}}  \bibinfo{volume}{30} (\bibinfo{year}{2017}),
  \bibinfo{pages}{5998--6008}.
\newblock


\bibitem[Veli{\v{c}}kovi{\'c} et~al\mbox{.}(2018)]%
        {velivckovic2018graph}
\bibfield{author}{\bibinfo{person}{Petar Veli{\v{c}}kovi{\'c}},
  \bibinfo{person}{Guillem Cucurull}, \bibinfo{person}{Arantxa Casanova},
  \bibinfo{person}{Adriana Romero}, \bibinfo{person}{Pietro Li{\`o}}, {and}
  \bibinfo{person}{Yoshua Bengio}.} \bibinfo{year}{2018}\natexlab{}.
\newblock \showarticletitle{Graph Attention Networks}. In
  \bibinfo{booktitle}{\emph{International Conference on Learning
  Representations}}.
\newblock


\bibitem[Wan et~al\mbox{.}(2018)]%
        {wan2018improving}
\bibfield{author}{\bibinfo{person}{Yao Wan}, \bibinfo{person}{Zhou Zhao},
  \bibinfo{person}{Min Yang}, \bibinfo{person}{Guandong Xu},
  \bibinfo{person}{Haochao Ying}, \bibinfo{person}{Jian Wu}, {and}
  \bibinfo{person}{Philip~S Yu}.} \bibinfo{year}{2018}\natexlab{}.
\newblock \showarticletitle{Improving automatic source code summarization via
  deep reinforcement learning}. In \bibinfo{booktitle}{\emph{Proceedings of the
  33rd ACM/IEEE International Conference on Automated Software Engineering}}.
  \bibinfo{pages}{397--407}.
\newblock


\bibitem[Wang et~al\mbox{.}(2021)]%
        {wang2021codet5}
\bibfield{author}{\bibinfo{person}{Yue Wang}, \bibinfo{person}{Weishi Wang},
  \bibinfo{person}{Shafiq Joty}, {and} \bibinfo{person}{Steven~CH Hoi}.}
  \bibinfo{year}{2021}\natexlab{}.
\newblock \showarticletitle{CodeT5: Identifier-aware Unified Pre-trained
  Encoder-Decoder Models for Code Understanding and Generation}. In
  \bibinfo{booktitle}{\emph{Proceedings of the 2021 Conference on Empirical
  Methods in Natural Language Processing}}. \bibinfo{pages}{8696--8708}.
\newblock


\bibitem[Wei et~al\mbox{.}(2019)]%
        {wei2019code}
\bibfield{author}{\bibinfo{person}{Bolin Wei}, \bibinfo{person}{Ge Li},
  \bibinfo{person}{Xin Xia}, \bibinfo{person}{Zhiyi Fu}, {and}
  \bibinfo{person}{Zhi Jin}.} \bibinfo{year}{2019}\natexlab{}.
\newblock \showarticletitle{Code generation as a dual task of code
  summarization}. In \bibinfo{booktitle}{\emph{Advances in Neural Information
  Processing Systems 2019}}. Neural Information Processing Systems (NIPS).
\newblock


\bibitem[Wei et~al\mbox{.}(2020)]%
        {wei2020retrieve}
\bibfield{author}{\bibinfo{person}{Bolin Wei}, \bibinfo{person}{Yongmin Li},
  \bibinfo{person}{Ge Li}, \bibinfo{person}{Xin Xia}, {and}
  \bibinfo{person}{Zhi Jin}.} \bibinfo{year}{2020}\natexlab{}.
\newblock \showarticletitle{Retrieve and refine: exemplar-based neural comment
  generation}. In \bibinfo{booktitle}{\emph{2020 35th IEEE/ACM International
  Conference on Automated Software Engineering (ASE)}}. IEEE,
  \bibinfo{pages}{349--360}.
\newblock


\bibitem[Wu et~al\mbox{.}(2021)]%
        {wu2021code}
\bibfield{author}{\bibinfo{person}{Hongqiu Wu}, \bibinfo{person}{Hai Zhao},
  {and} \bibinfo{person}{Min Zhang}.} \bibinfo{year}{2021}\natexlab{}.
\newblock \showarticletitle{Code Summarization with Structure-induced
  Transformer}. In \bibinfo{booktitle}{\emph{Findings of the Association for
  Computational Linguistics: ACL-IJCNLP 2021}}. \bibinfo{pages}{1078--1090}.
\newblock


\bibitem[Xia et~al\mbox{.}(2017)]%
        {xia2017measuring}
\bibfield{author}{\bibinfo{person}{Xin Xia}, \bibinfo{person}{Lingfeng Bao},
  \bibinfo{person}{David Lo}, \bibinfo{person}{Zhenchang Xing},
  \bibinfo{person}{Ahmed~E Hassan}, {and} \bibinfo{person}{Shanping Li}.}
  \bibinfo{year}{2017}\natexlab{}.
\newblock \showarticletitle{Measuring program comprehension: A large-scale
  field study with professionals}.
\newblock \bibinfo{journal}{\emph{IEEE Transactions on Software Engineering}}
  \bibinfo{volume}{44}, \bibinfo{number}{10} (\bibinfo{year}{2017}),
  \bibinfo{pages}{951--976}.
\newblock


\bibitem[Yang et~al\mbox{.}(2021)]%
        {yang2021incbl}
\bibfield{author}{\bibinfo{person}{Zhou Yang}, \bibinfo{person}{Jieke Shi},
  \bibinfo{person}{Shaowei Wang}, {and} \bibinfo{person}{David Lo}.}
  \bibinfo{year}{2021}\natexlab{}.
\newblock \showarticletitle{Incbl: Incremental bug localization}. In
  \bibinfo{booktitle}{\emph{2021 36th IEEE/ACM International Conference on
  Automated Software Engineering (ASE)}}. IEEE, \bibinfo{pages}{1223--1226}.
\newblock


\bibitem[Zhang et~al\mbox{.}(2020)]%
        {zhang2020retrieval}
\bibfield{author}{\bibinfo{person}{Jian Zhang}, \bibinfo{person}{Xu Wang},
  \bibinfo{person}{Hongyu Zhang}, \bibinfo{person}{Hailong Sun}, {and}
  \bibinfo{person}{Xudong Liu}.} \bibinfo{year}{2020}\natexlab{}.
\newblock \showarticletitle{Retrieval-based neural source code summarization}.
  In \bibinfo{booktitle}{\emph{2020 IEEE/ACM 42nd International Conference on
  Software Engineering (ICSE)}}. IEEE, \bibinfo{pages}{1385--1397}.
\newblock


\bibitem[Zhang et~al\mbox{.}(2019)]%
        {zhang2019novel}
\bibfield{author}{\bibinfo{person}{Jian Zhang}, \bibinfo{person}{Xu Wang},
  \bibinfo{person}{Hongyu Zhang}, \bibinfo{person}{Hailong Sun},
  \bibinfo{person}{Kaixuan Wang}, {and} \bibinfo{person}{Xudong Liu}.}
  \bibinfo{year}{2019}\natexlab{}.
\newblock \showarticletitle{A novel neural source code representation based on
  abstract syntax tree}. In \bibinfo{booktitle}{\emph{2019 IEEE/ACM 41st
  International Conference on Software Engineering (ICSE)}}. IEEE,
  \bibinfo{pages}{783--794}.
\newblock


\end{thebibliography}


\end{document}